\documentclass{aa} \usepackage{graphicx}
\usepackage{txfonts}
\usepackage{natbib}

\bibpunct{(}{)}{;}{a}{}{,}

\begin{document}

\title{On the origin of H$_2$CO abundance enhancements in low-mass
protostars}


\author{F. L. Sch\"oier\inst{1,2}  \and
             J. K. J{\o}rgensen\inst{1}
            E. F. van Dishoeck\inst{1} \and
            G. A. Blake\inst{3} }

\offprints{F. L. Sch\"oier \\ \email{fredrik@astro.su.se}}

\institute{Leiden Observatory, P.O. Box 9513, NL-2300 RA Leiden, The
Netherlands \and Stockholm Observatory, AlbaNova, SE-106 91 Stockholm,
Sweden \and Division of Geological and Planetary Sciences, California
Institute of Technology, MS 150-21, Pasadena, CA 91125, USA}

\date{Received; accepted}

\abstract{High angular resolution H$_2$CO 218~GHz line observations
have been carried out toward the low-mass protostars
\object{IRAS~16293--2422} and \object{L1448--C} using the Owens Valley
Millimeter Array at $\sim$2$\arcsec$ resolution.  Simultaneous 1.37~mm
continuum data reveal extended emission which is compared with that
predicted by model envelopes constrained from single-dish data.  For
\object{L1448--C} the model density structure works well down to the
400 AU scale to which the interferometer is sensitive. For
\object{IRAS 16293--2422}, a known proto-binary object, the
interferometer observations indicate that the binary has cleared much
of the material in the inner part of the envelope, out to the binary
separation of $\sim$800~AU.  For both sources there is excess
unresolved compact emission centered on the sources, most likely due
to accretion disks $\lesssim$200\,AU in size with masses of 
$\gtrsim$0.02 M$_{\odot}$ (L1448--C) and $\gtrsim$0.1~M$_{\odot}$
(IRAS~16293--2422).  The H$_2$CO data for both sources are dominated
by emission from gas close to the positions of the continuum peaks.
The morphology and velocity structure of the H$_2$CO array data have
been used to investigate whether the abundance enhancements inferred
from single-dish modelling are due to thermal evaporation of ices or
due to liberation of the ice mantles by shocks in the inner envelope.
For \object{IRAS 16293--2422} the H$_2$CO interferometer observations
indicate the presence of large scale rotation roughly perpendicular to
the large scale CO outflow.  The H$_2$CO distribution differs from
that of C$^{18}$O, with C$^{18}$O emission peaking near MM1 and
H$_2$CO stronger near MM2. For \object{L1448--C}, the region of
enhanced H$_2$CO emission extends over a much larger scale $>$1$''$ than
the radius of $50-100$~K ($0\farcs6-0\farcs15$) where thermal
evaporation can occur. The red-blue asymmetry of the emission is
consistent with the outflow; however the velocities are significantly
lower.  The H$_2$CO $3_{22}-2_{21}/3_{03}-2_{02}$ flux ratio derived
from the interferometer data is significantly higher than that found
from single-dish observations for both objects, suggesting that the
compact emission arises from warmer gas.  Detailed radiative transfer
modeling shows, however, that the ratio is affected by abundance
gradients and optical depth in the $3_{03}-2_{02}$ line.  It is
concluded that a constant H$_2$CO abundance throughout the envelope
cannot fit the interferometer data of the two H$_2$CO lines
simultaneously on the longest and shortest baselines.  A scenario in
which the H$_2$CO abundance drops in the cold dense part of the
envelope where CO is frozen out but is undepleted in the outermost
region provides good fits to the single-dish and interferometer data
on short baselines for both sources. Emission on the longer baselines
is best reproduced if the H$_2$CO abundance is increased by about an order
of magnitude from  $\sim$$10^{-10}$ to $\sim$$10^{-9}$ 
in the inner parts of the envelope  due to
thermal evaporation when the temperature exceeds $\sim$50~K.  The
presence of additional H$_2$CO abundance jumps in the innermost hot
core region or in the disk cannot be firmly
established, however, with the present sensitivity and resolution.
Other scenarios, including weak outflow-envelope
interactions and photon heating of the envelope, are discussed and
predictions for future generation interferometers are presented,
illustrating their potential in distinguishing these competing
scenarios.   }

\maketitle

\section{Introduction}
Recent observational studies have shown that the inner ($<$ few
hundred AU) envelopes of low-mass protostars are dense 
($\gtrsim$10$^6$~cm$^{-3}$) and warm ($\gtrsim$80~K) \citep{Blake94, Ceccarelli00a,
Joergensen02, Schoeier02, Shirley02}, as expected from scaling of high-mass
protostars \citep{Ceccarelli96, Ivezic97}. In high-mass objects, these
warm and dense regions are known to have a rich chemistry with high
abundances of organic molecules due to the thermal evaporation of ices
\citep[e.g.,][]{Blake87, Charnley92}.  Detailed modeling of
multi-transition single-dish lines toward the deeply embedded low-mass
protostar \object{IRAS 16293--2422} has demonstrated that similar
enhancements of molecules like H$_2$CO and CH$_3$OH can occur for
low-mass objects \citep{Dishoeck95, Ceccarelli00b,Schoeier02}.
Recently, \citet{Maret04} have suggested that this is a common
phenomenon in low-mass protostars. The location at which this
enhancement occurs is consistent with the radius at which ices are
expected to thermally evaporate off the grains ($T\gtrsim 90$~K).
Moreover, large organic molecules have recently been detected toward
\object{IRAS 16293--2422} \citep{Cazaux03}, showing that low-mass hot
cores may have a similar chemical complexity as the high-mass
counterparts in spite of their much shorter timescales
\citep{Schoeier02}.

Alternatively, shocks due to the interaction of the outflow with the
inner envelope can liberate grain mantle material over a larger area
than can thermal heating. Additionally, the bipolar outflow will
excavate a biconical cavity in the envelope through which UV- and
X-ray photons can escape. The back scattering of such photons into the
envelope by low-density dust in the cavity can significantly heat the
envelope surrounding the cavity \citep[e.g.,][]{Spaans95}. This would
produce regions of warm gas ($\sim$100~K) in the envelope on
much larger scales than otherwise possible.  High angular resolution
observations are needed to pinpoint the origin of the abundance
enhancements and distinguish between these various scenarios.

We present here observations of H$_2$CO toward two low-mass
protostars, \object{IRAS 16293--2422} and \object{L1448--C} (also
known as \object{L1448--mm}), at 218~GHz (1.4~mm) using the Owens
Valley Radio Observatory (OVRO) Millimeter Array at $\sim$2$\arcsec$
resolution.  The frequency setting includes the
$3_{03}\rightarrow2_{02}$ and $3_{22}\rightarrow2_{21}$ H$_2$CO lines,
whose ratio is a measure of the gas temperature of the circumstellar
material. Both \object{IRAS 16293--2422} and \object{L1448--C} are
deeply embedded class 0 protostars \citep{andre93} which drive large
scale ($\sim$\,arcmin) bipolar outflows \citep{Walker88, Mizuno90,
Bachiller90, Bachiller91,Stark04}.  For other low-mass objects,
molecules such as SiO are clearly associated with the outflow (e.g.,
L1448: Guilloteau et al.\ 1992\nocite{Guilloteau92}; NGC 1333 IRAS4:
Blake 1995\nocite{Blake95}), whereas optically thick lines from other species such as
HCO$^+$ and HCN are found to `coat' the outflow walls (e.g., B5 IRS1:
Langer et al.\ 1996\nocite{Langer96}; L1527 and Serpens SMM1:
Hogerheijde et al.\ 1997\nocite{Hogerheijde97},
1999\nocite{Hogerheijde99}).  The extent of this emission can be
larger than $10\arcsec$, which should be readily distinguishable from
the $\sim$1$\arcsec$ hot inner envelope with current interferometers.

Previous millimeter aperture synthesis observations of \object{IRAS
16293--2422} have revealed two compact components coincident with
radio continuum emission, indicative of a protobinary source
\citep{Mundy90, Mundy92}.  The line emission of 10 molecular species
at $\sim$5$\arcsec$ resolution reveals that there is a red-blue
asymmetry indicative of rotation perpendicular to the outflow
direction (Sch\"oier et al., in prep.). The morphology of the emission
picked up by the interferometer suggests that it may be produced in
regions of compressed gas as a result of interaction between the
outflow and the envelope. Previous data on \object{L1448--C} show a
compact continuum source at millimetre wavelengths and that SiO is a
good tracer of the large velocity outflow associated with this source
\citep{Guilloteau92}.

Since most of the extended emission is resolved out by the
interferometer, a good physical and chemical model of the envelope is
a prerequisite for a thorough interpretation of the aperture synthesis
data. In recent years, much progress has been made in obtaining
reliable descriptions of the density and temperature structures in the
dusty envelopes around young stellar objects, based on thermal
continuum emission \citep{Chandler00, Hogerheijde00b, Motte01,
Joergensen02, Schoeier02, Shirley02}.  The physical structures of
\object{IRAS 16293--2422} and \object{L1448--C} have recently been
derived from single-dish continuum observations, with the results
summarized in Table~\ref{parameters} \citep{Joergensen02,Schoeier02}.

In \S\ref{dustmodelling}, we first test the validity of these envelope
models at the small scales sampled by the continuum interferometer
data. In \S\ref{h2co}, the H$_2$CO results are presented and
analyzed. For \object{IRAS 16293--2422}, C$^{18}$O observations
are also available. This is followed by a discussion on the origin of
H$_2$CO and estimates for what future generation telescopes might  
reveal in
\S\ref{discussion} and by conclusions in \S\ref{conclusions}.

\begin{table}
\caption[]{Source and envelope parameters for \object{IRAS 16293--2422}  
and  \object{L1448--C}}
  \label{parameters}
$$
\begin{array}{p{0.61\linewidth}cc}
\hline\hline
\noalign{\smallskip}
  & \mathrm{IRAS\ 16293}^a & \mathrm{L1448}^b \\
\hline
\noalign{\smallskip}
Distance, $D$ (pc) & 160 & 220 \\
Luminosity, $L$ (L$_{\sun}$) & 27 & 5\\
Inner radius, $r_{\mathrm{i}}$ (AU) & 32.1 & 9.0\\
Outer radius, $r_{\mathrm{e}}$ (10$^4$\,AU) & 0.80 & 0.81 \\
Density power law index, $\alpha$ & 1.7 & 1.4 \\
Density at 1000\,AU, $n_0$(H$_2$) ($10^{6}$\,cm$^{-3}$) & 6.7& 0.75\\
Column density$^c$, $N(\mathrm{H}_2)$ (10$^{24}$\,cm$^{-2}$)
  & 1.6 & 0.17\\
Envelope mass$^c$, $M_{\mathrm{env}}$ (M$_{\sun}$) & 5.4 & 0.93 \\
\hline
\noalign {\smallskip}
$^a$ From \citet{Schoeier02} \\
$^b$ From \citet{Joergensen02} \\
$^c$ Within the outer radius $r_{\mathrm{e}}$\\
\end{array}
$$
\end{table}

\section{Observations and data reduction}
\subsection{Interferometer data}
The two protostars \object{IRAS 16293--2422}
($\alpha_{2000}=16^{\mathrm h}32^{\mathrm m}22\fs8$,
$\delta_{2000}=-24{\degr}28{\arcmin}33\farcs0$) and \object{L1448--C}
($\alpha_{2000}=3^{\mathrm h}25^{\mathrm m}38\fs8$,
$\delta_{2000}=30{\degr}44{\arcmin}05\farcs0$) were observed with the
Owens Valley Radio Observatory (OVRO) Millimeter
Array\footnote{Research with the Owens Valley Millimeter Array,
operated by California Institute of Technology, is supported by NSF
grant AST 99-81546.} between September 2000 and March 2002.  The
H$_2$CO $3_{03}\rightarrow 2_{02}$ and $3_{22}\rightarrow 2_{21}$ line
emission at 218.222 and 218.475\,GHz, respectively, was obtained
simultaneously with the continuum emission at 1.37\,mm.  \object{IRAS
16293--2422} was observed in the L and E configurations, while
\object{L1448--C} was observed in the C, L and H configurations,
corresponding to projected baselines of $8-80$ and $8-120$~k$\lambda$,
respectively. The complex gains were calibrated by regular
observations of the quasars \object{NRAO 530} and \object{1622--253}
for \object{IRAS 16293--2422} and \object{0234$+$285} for
\object{L1448--C}, while flux calibration was done using observations
of Uranus and Neptune for each track, both using the MMA package
developed for OVRO data by \citet{Scoville93}.  The subsequent
data-reduction and analysis was performed using MIRIAD
\citep{Sault95}.

Further reduction of the data was carried out using the standard
approach by flagging clearly deviating phases and amplitudes.
The continuum data were self-calibrated and the resulting
phase corrections were applied to the spectral line data, optimizing
the signal-to-noise. The natural-weighted continuum observations for
\object{IRAS 16293--2422} and \object{L1448--C} have typical 1$\sigma$
noise levels better than 20 and 3\,mJy\,beam$^{-1}$ with beam sizes of
3$\farcs$9$\times$1$\farcs$9 and 2$\farcs$6$\times$2$\farcs$3,
respectively. The relatively high noise levels for \object{IRAS
16293--2422} reflect the low elevation at which this source is
observable from Owens Valley, which both increases the system
temperatures and decreases the time available per transit.
The data
were then CLEANed \citep{Hoegbom74} down to the 2$\sigma$ noise level.

For \object{IRAS 16293--2422} additional archival C$^{18}$O
$J=2\rightarrow 1$ line data obtained in 1993 using the OVRO array
are presented, which were reduced in the same way as described above.

\subsection{Single-dish data}
It is well-known that the millimeter aperture synthesis observations
lack sensitivity to extended emission due to discrete sampling in the
$(u,v)$ plane and, in particular, missing short-spacings. In order to  
quantify
this single-dish observations were performed using the James Clerk
Maxwell Telescope (JCMT)\footnote{The JCMT is operated by the Joint
Astronomy Centre in Hilo, Hawaii on behalf of the Particle Physics and
Astronomy Research Council in the United Kingdom, the National
Research Council of Canada and the Netherlands Organization for
Scientific Research.}. The continuum data are taken largely from the
JCMT archive\footnote{\tt http://www.jach.hawaii.edu/JACpublic/JCMT/}
and have been presented in \citet{Schoeier02} and
\citet{Joergensen02}. For H$_2$CO, a 25-point grid centered on the
adopted source position and sampled at 10$\arcsec$ spacing was
obtained in September 2002 for \object{IRAS 16293--2422}, with both
H$_2$CO 218 GHz lines covered in a single spectral setting.  The
observations were obtained in a beam-switching mode using a
$180\arcsec$ chop throw.  The data were calibrated using the
chopper-wheel method and the resulting antenna temperature was
converted into main-beam brightness temperature, $T_{\mathrm{mb}}$,
using the main-beam efficiency $\eta_{\mathrm{mb}}=0.69$.
For \object{L1448--C}, a single spectrum at the source position was  
taken
that includes both transitions.

\begin{figure}
\centering{\includegraphics[width=8cm]{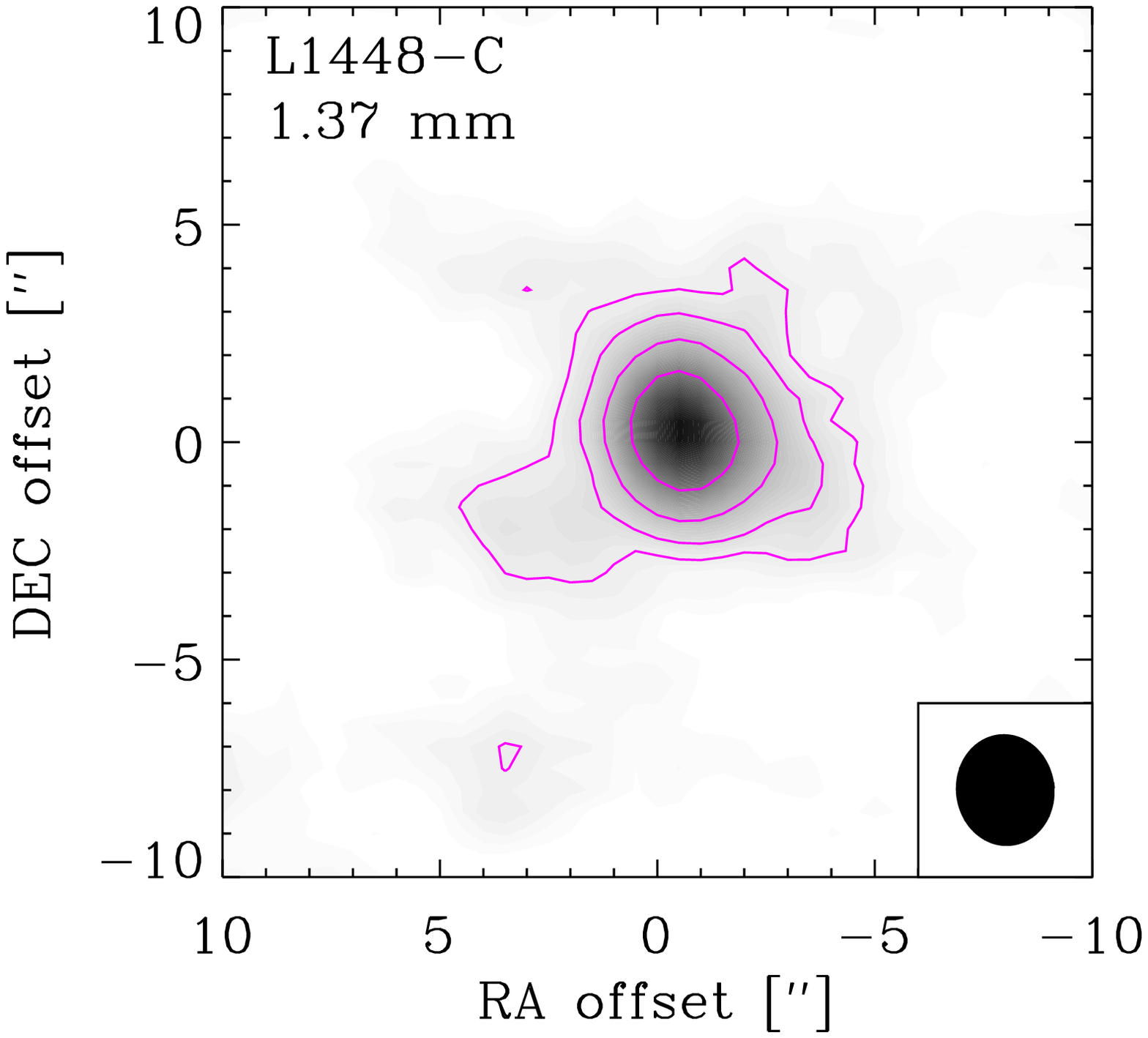}
\caption{OVRO interferometer maps of the 1.37~mm  continuum emission
toward \object{L1448--C}. Contours start at
9~mJy\,beam$^{-1}$ and the first contour corresponds to the
`2$\sigma$'-level as estimated from the maps. Each successive contour
is a multiple of two of the preceding value.  The peak value is
0.13~Jy\,beam$^{-1}$ and the synthesized beam in this and subsequent  
images
is depicted by the filled ellipse in the lower-right corner.}
\label{cont_l1448}}
\end{figure}
\section{Continuum emission: disk and envelope structure}
\label{dustmodelling}
\subsection{L1448--C}
\label{l1448_dust}

In Fig.~\ref{cont_l1448}, the 221.7\,GHz (1.37\,mm) continuum emission
toward \object{L1448--C} is presented.  Only a single compact
component is seen, with faint extended emission.  The total continuum
flux density at 1.37~mm observed with OVRO is 0.32~Jy, only 35\% of
the flux observed by \citet{Motte01} (0.9~Jy) at 1.3~mm using the IRAM
30~m telescope. The compact component, located at ($-0\farcs 6$,
$0\farcs 2$) from the pointing centre, has been fitted with a Gaussian
in the ($u,v$) plane, resulting in an estimated size of
1$\farcs$0$\times$0$\farcs$6 and an upper limit to the diameter of
$\sim$170~AU for the adopted distance of 220~pc.

\begin{figure}
\centering{\includegraphics[width=8cm]{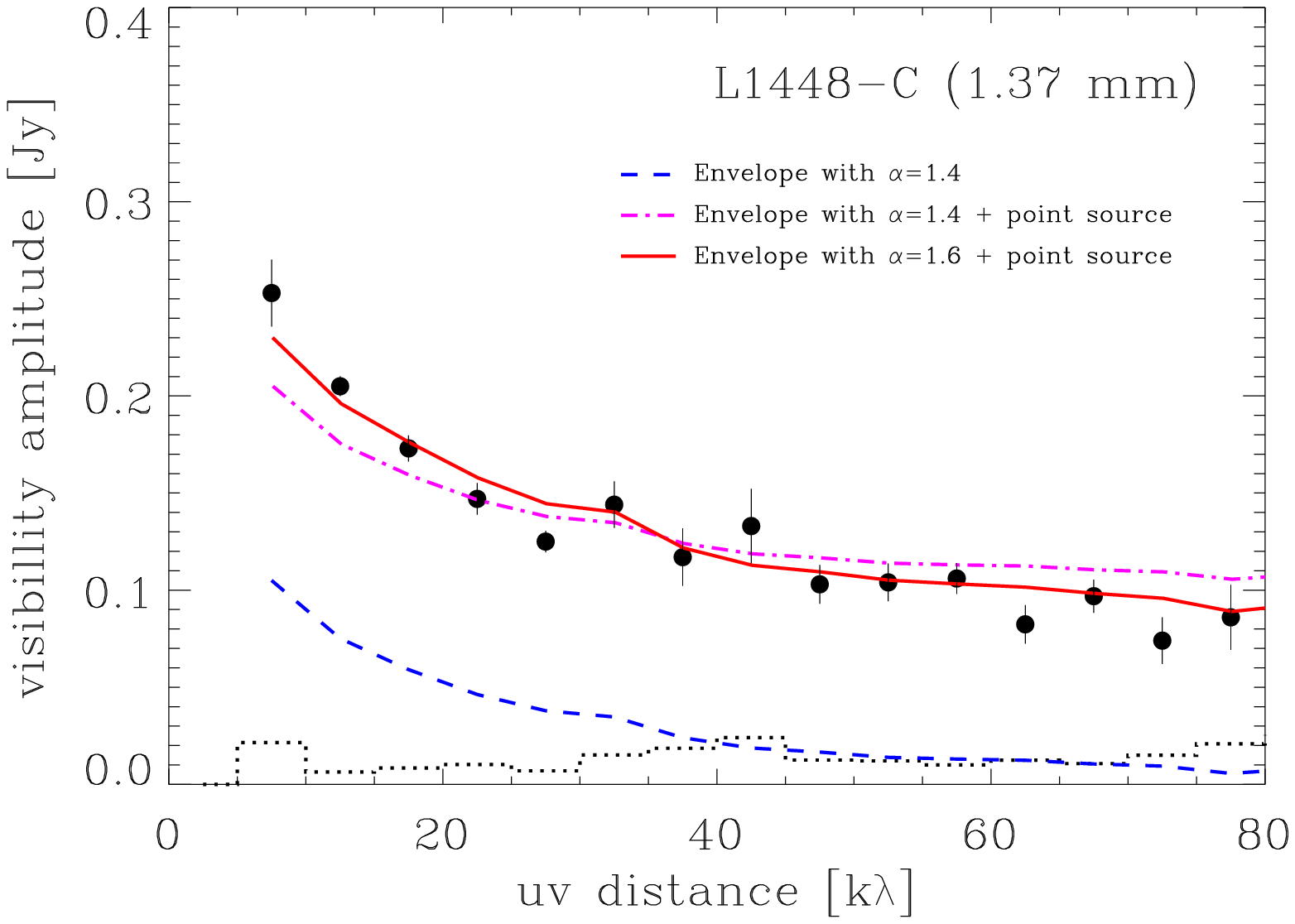}
\caption{Visibility amplitudes of the observed 1.37~mm continuum
emission obtained at OVRO towards \object{L1448--C} as functions of
the projected baseline length, binned to 5~k$\lambda$, from the phase
center, taken to be at ($-0\farcs 6,0\farcs 2$). The observations are
plotted as filled symbols with 1$\sigma$ error bars. The dotted histogram represents the 
zero-expectation level.
Also shown are
predictions based on a realistic physical model for the source
\citep{Joergensen02}, with the same ($u,v$) sampling as the
observations (see text for details). Unresolved compact emission,
presumably from a circumstellar disk, must be added to that
from the envelope to produce an acceptable fit.}
\label{cont_vis_l1448}}
\end{figure}

\citet{Joergensen02} determined the actual temperature and density
distribution of the circumstellar envelope of \object{L1448--C} from
detailed modeling of the observed continuum emission (see
Table~\ref{parameters}). In addition to the spectral energy
distribution (SED), resolved images at 450 and 850~$\mu$m obtained
with the SCUBA bolometer array at the JCMT were used to constrain the
large scale envelope structure. The interferometer data constrain the  
envelope
structure at smaller scales ($\sim$2$\arcsec$) than the JCMT
single-dish data ($\sim$10-20$\arcsec$).
In order to investigate whether this
envelope model can be reconciled with the flux picked up by the
interferometer, the same ($u,v$) sampling was applied to the predicted
brightness distribution at 1.37~mm from the model
envelope.

Fig.~\ref{cont_vis_l1448} compares the observed visibility amplitudes
of \object{L1448--C} with the model predictions. First, it is clear
that the envelope alone cannot reproduce the data, but that an
unresolved compact source, presumably the disk, needs to be added.
The combination of this point source and the best fit envelope model
of \citet{Joergensen02} with a density structure falling off as
$r^{-\alpha}$, where $\alpha=1.4$, dramatically improves the fit to
the visibilities in the $(u,v)$ plane.  A slightly steeper density
structure of $\alpha=1.6$ is preferred by the interferometer
data. Given the mutual uncertainties of $\pm 0.2$ in $\alpha$ these
results are still in good agreement with each other and with those of
\citet{Shirley02}, who found a slope $\alpha=1.7$ in their analysis
for a somewhat larger ($r_e=45000$~AU) envelope. The remaining point
source flux for \object{L1448--C} is estimated to be 100 and 75~mJy
when $\alpha$ is taken to be 1.4 and 1.6, respectively. This is $\sim
10$\% of the total single-dish source flux. The impact of changing
different envelope parameters such as density slope and location of
inner envelope radius was tested in \citet{Joergensen04a}, for the embedded 
low-mass protostar \object{NGC 1333--IRAS2A}, and found to
lead to an uncertainty of about $\pm 25$\% on the derived point source
flux.

\begin{figure}
\centering{\includegraphics[width=7.5cm, angle=-90]{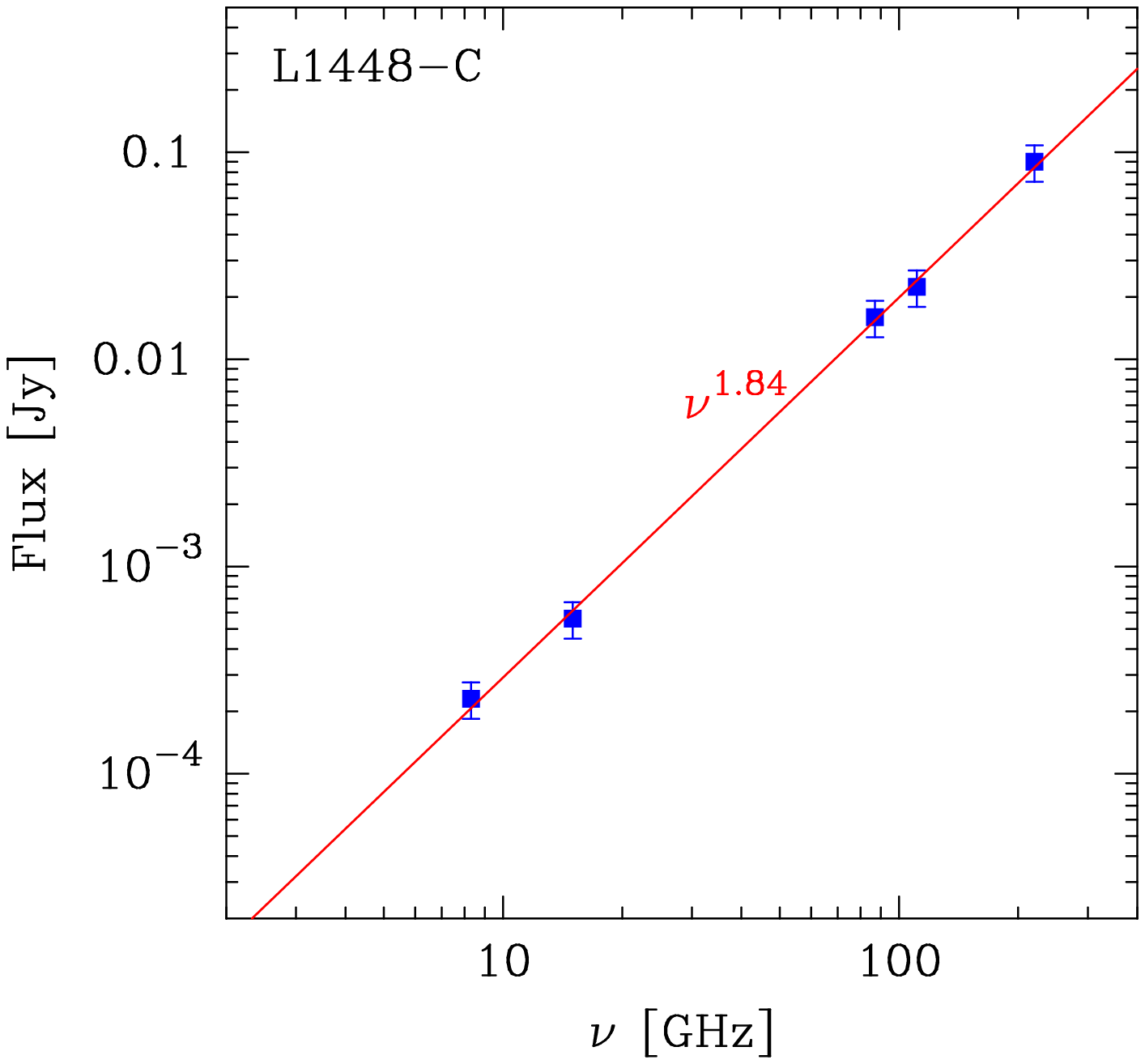}
\caption{Continuum flux observations of the compact emission toward
\object{L1448--C} (squares with error bars). The solid line shows a
fit to the data using $F\propto\nu^\beta$, where $\beta=1.84\pm 0.08$ is the
spectral index.}
\label{alpha_l1448}}
\end{figure}

The point source flux estimated here at 1.37~mm agrees very well with
the spectral index of $1.84\pm 0.08$ found from other
mm and cm wavelength observations \citep{Curiel90, Guilloteau92,
Looney00, Reipurth02} as shown in Fig.~\ref{alpha_l1448}. The quoted 
standard deviation is based on assuming a 20\% uncertainty in all the fluxes. Similarly,
\citet{Joergensen04a} found a spectral index of 1.9 for the point source
associated with the low-mass
protostar \object{NGC 1333--IRAS2A}.  The spectral index is consistent
with optically thick thermal emission and its favoured origin is that
from an unresolved accretion disk.


%

Assuming the point source emission to be thermal the mass of the
compact region can be estimated from
\begin{equation}
\label{diskmass}
M=\frac{F_{\nu}\Psi D^2}{\kappa_{\nu}B_{\nu}(T_{\mathrm d})} \left(  
\frac{\tau_{\nu}}{1-e^{-\tau_{\nu}}} \right),
\end{equation}
where $F_{\nu}$ is the flux, $\Psi$ is the gas-to-dust ratio (assumed
to be equal to 100), $D$ is the distance, $\kappa_{\nu}$ is the dust
opacity, $B_{\nu}$ is the Planck function at a characteristic dust
temperature $T_{\mathrm d}$ and $\tau_{\nu}$ is the optical depth.
The adopted dust opacity at 1.37~mm, 0.8~cm$^2$\,g$^{-1}$, is
extrapolated from the opacities presented by \citet{Ossenkopf94} for
grains with thin ice mantles. These opacities were used also in the
radiative transfer analysis of the envelope.  For a dust temperature
in the range $100-40$~K, the estimated disk mass in the optically
thin limit is $0.016-0.042$~M$_{\odot}$ when a point source flux of
75~mJy is used.  This should be treated as a lower limit since the
emission is likely to be optically thick at 1.37~mm, as suggested by
the spectral index.

It is difficult to estimate accurate disk masses for deeply embedded
sources since it involves a good knowledge about the envelope
structure, in addition to, e.g., the disk temperature. For more
evolved protostars where the confusion with the envelope is less
problematic, disk masses of $\sim$0.01 - 0.08~M$_{\odot}$ are derived
\citep[e.g.,][]{Looney00, Mundy00}, comparable to the values found here.

\begin{figure}
\centering{\includegraphics[width=8cm]{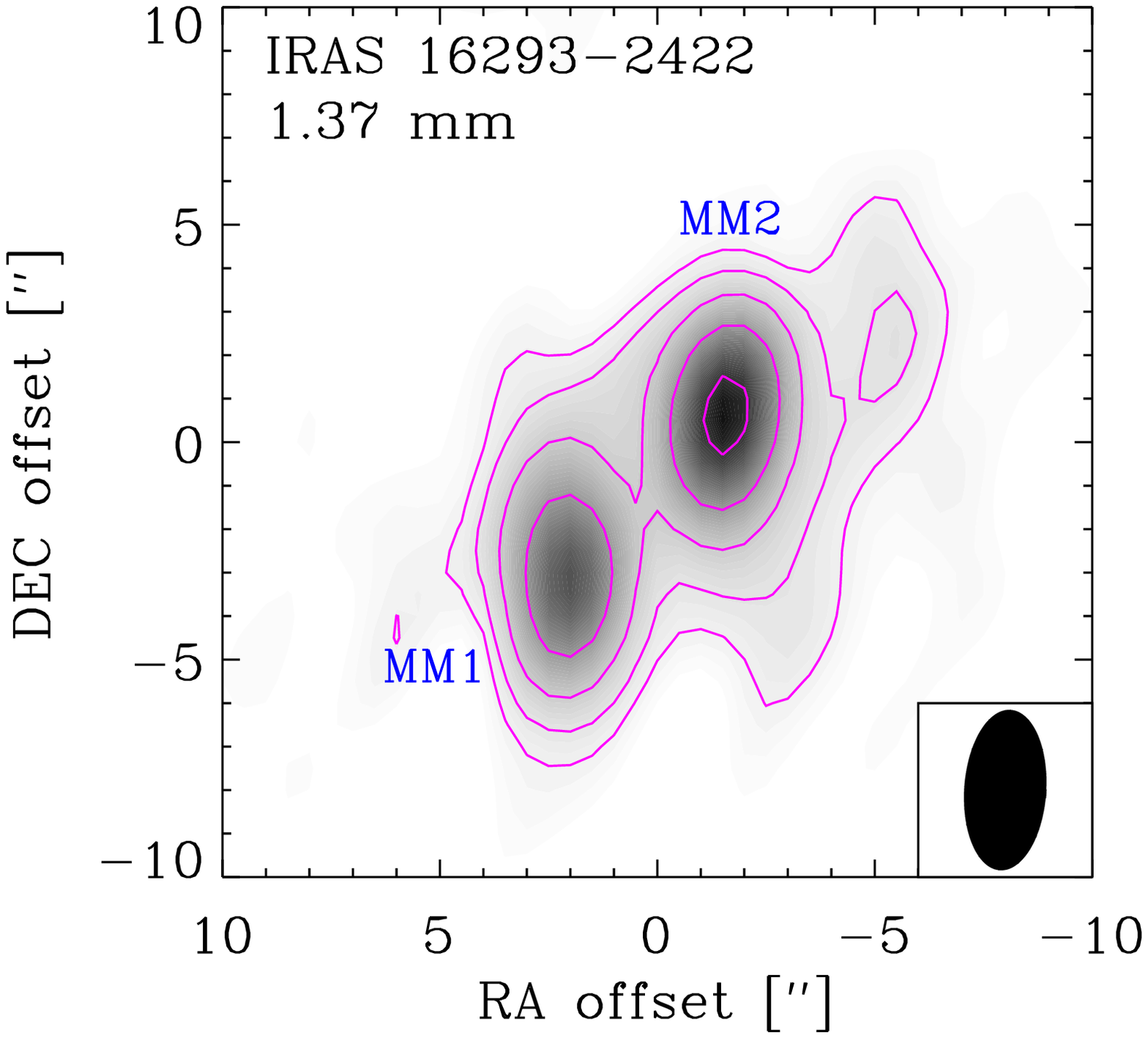}
\caption{OVRO interferometer maps of the continuum emission at
1.37\,mm towards \object{IRAS 16293--2422}. Contours start at
60\,mJy\,beam$^{-1}$ and the first contour corresponds to the
`2$\sigma$'-level as estimated from the maps. Each successive contour
is a multiple of two of the preceding value.  The emission peaks at
0.82 and 1.1~Jy\,beam$^{-1}$ at MM1 and MM2, respectively. }
\label{cont_iras}}
\end{figure}

\begin{figure}
\centering{\includegraphics[width=8cm]{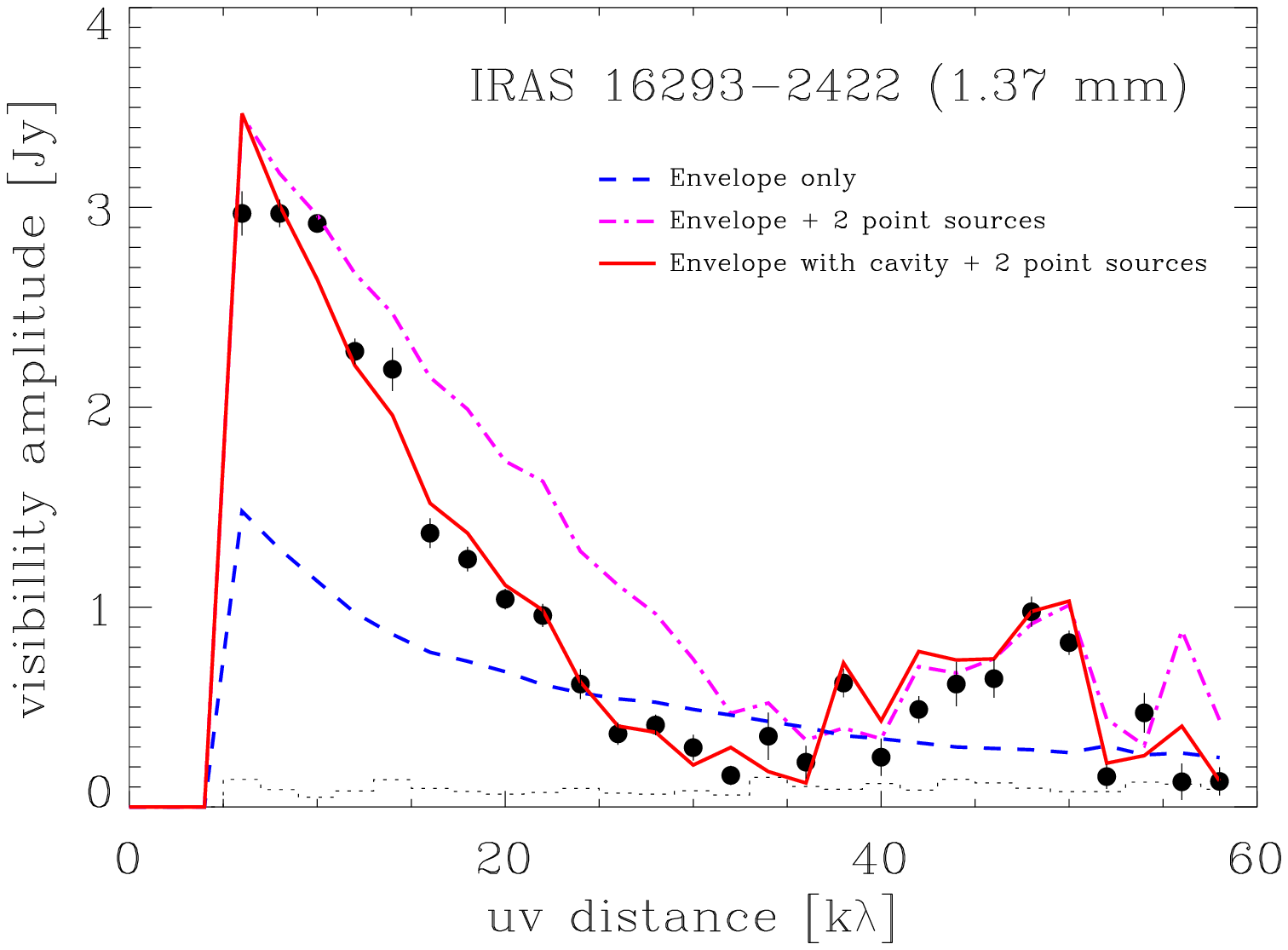}
\caption{Visibility amplitudes of the observed 1.37~mm emission
obtained at OVRO towards \object{IRAS 16293--2422} as functions of the
projected baseline length, binned to 2~k$\lambda$, from the phase
center, taken to be at (0,0). The observations are plotted as filled
symbols with 1$\sigma$ error bars. The dotted histogram represents the 
zero-expectation level.
Also shown are predictions based on
a realistic physical model for the source \citep{Schoeier02}, with the
same ($u,v$) sampling as the observations (see text for
details). Unresolved compact emission, presumably from two
circumstellar disks, needs to be added to that emanating from the
envelope in order to obtain an acceptable fit. A model envelope
with a cavity (solid line), in addition to the unresolved compact
emission, is shown to best reproduce the observations.}
\label{cont_vis_iras}}
\end{figure}
\begin{figure}
\centering{\includegraphics[width=8cm]{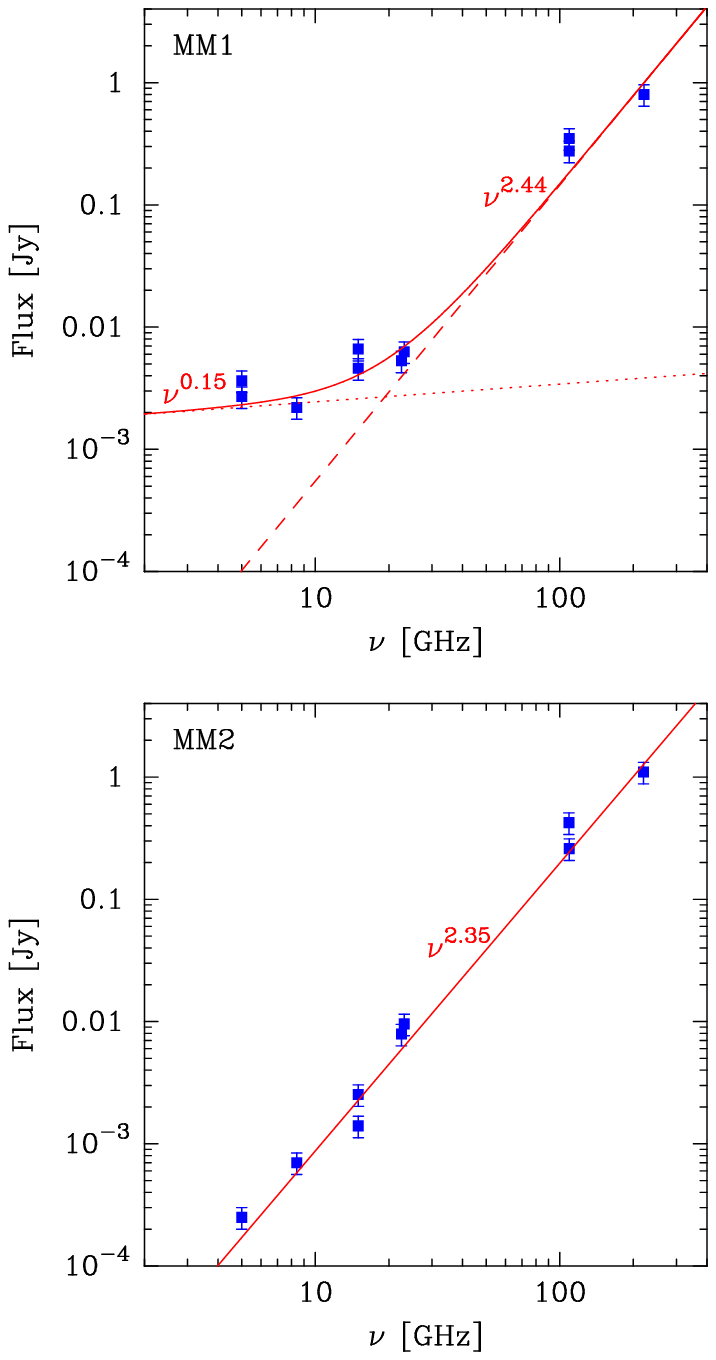}
\caption{Continuum flux observations of the compact emission towards
\object{IRAS 16293--2422} (squares with error bars). The solid lines
show fits to the data using $F\propto\nu^\beta$, where $\beta$
is the spectral index. For MM1 a combination of two powerlaws were used.}
\label{alpha_iras}}
\end{figure}

\subsection{IRAS 16293--2422}
\label{dust_iras}

For the proto-binary object \object{IRAS 16293--2422}, two unresolved
continuum sources are detected separated by approximately $5\arcsec$
(Fig.~\ref{cont_iras}).
The total observed continuum flux density at
1.37~mm is about 3.5~Jy. This is roughly 50\% of the flux obtained
from mapping with single-dish telescopes \citep{Walker90, Andre94},
indicating that the interferometer resolves out some of the emission.
The positions of the continuum sources [($2\farcs 0$, $-2\farcs 9$)
and ($-1\farcs 6$, $0\farcs 5$)] are consistent with the two 3~mm
sources MM1 (southeast) and MM2 (northwest) found by
\citet{Mundy92}. At the distance of \object{IRAS 16293--2422} (160~pc)
the projected separation of the sources is about 800~AU.

\citet{Schoeier02} have modeled the circumbinary envelope of
\object{IRAS 16293--2422} in detail based on SCUBA images and the
measured SED. Fig.~\ref{cont_vis_iras} shows that this envelope alone
cannot fit the compact sources. The addition of two compact sources, 
at the locations of MM1 and MM2,  to
the best-fit envelope model of Sch\"oier et al.\ (2002) produces the
correct amount of flux at the longest baselines and the smaller
baselines, but now provides too much emission at intermediate
baselines ($10-30$~k$\lambda$, $\sim$10$\arcsec$), i.e., at scales of
the binary separation.

It is found that an envelope model which is void of material on scales
smaller than the binary separation best reproduces the observed
visibilities, in combination with the compact emission. For this
`cavity' model a slightly steeper density profile is obtained,
$\alpha=1.9$, from re-analyzing the SCUBA images and the SED.  Also,
the temperature is higher within $r\approx 1.5\times10^{16}$~cm
(1000~AU) compared to the standard envelope, although the temperature
never exceeds 80~K in the cavity model.

Theory has shown that an embedded binary system will undergo tidal
truncation and gradually clear its immediate environment due to
transfer of angular momentum from the binary to the disk.  Thus, an
inner gap or cavity with very low density is produced
\citep[e.g.,][]{Bate97, Gunther02}. Two binary sources in the T Tauri
stage have been imaged in great detail; \object{GG Tau} and \object{UY
Aur} \citep[e.g.,][]{Dutrey94, Duvert98}.  \citet{Wood99} estimate
that \object{GG Tau} has cleared its inner 200~AU radius of material
and that the bulk of material is located in a circumbinary ring of
thickness 600~AU.  \object{IRAS 16293--2422} could possibly be a `GG
Tau in the making'.

\begin{figure}
\centering{\includegraphics[width=8cm]{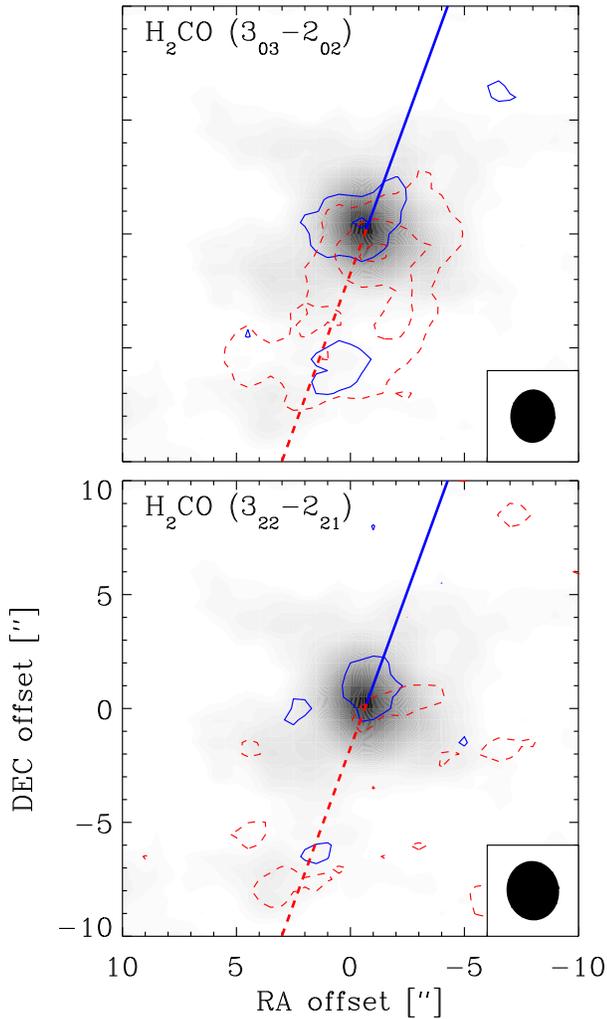}}
\caption{OVRO interferometer maps of H$_2$CO emission (contours)
overlayed on the 1.37\,mm continuum emission (greyscale) for
\object{L1448--C}.  The H$_2$CO emission has been separated into a red
(dashed lines) and a blue (solid lines) part (see text for
details). Contours start at at 0.2\,Jy\,beam$^{-1}$\,km\,s$^{-1}$
(2$\sigma$) and each succesive contour is a multiple of this
value. Also indicated are the directions of the large scale CO
outflow. }
\label{l1448_h2co}
\end{figure}

The emission from the two unresolved components is estimated to
contribute $\sim$25\% (1.8~Jy) to the total flux at 1.37~mm. In
Fig.~\ref{alpha_iras} flux estimates for the compact components around
MM1 and MM2 are compared to those at cm to mm wavelengths
\citep{Wootten89, Estalella91, Mundy92, Looney00}. The emission from
MM2 is well fitted over the entire region using a spectral index of
$2.35\pm 0.06$, consistent with thermal emission from an optically
thick disk. For MM1 a combination of two power laws provides the best
fit. At shorter wavelengths a spectral index of $2.44\pm 0.16$ is
observed, presumably thermal disk emission.  At longer wavelengths a
much lower index of $0.15\pm 0.15$ is found consistent with free-free
emission from an ionized stellar wind or jet. MM1 is indeed thought to
be the source driving the large scale outflow associated with
\object{IRAS 16293--2422}.  In comparison no active outflow is known
for MM2. However, it has been suggested that MM2 is responsible for a
fossilised flow in the E--W direction $\approx$10$\arcsec$ north of MM1
\citep[see][ and references therein]{Stark04}.

Gaussian fits to the sizes of these disks in the ($u,v$) plane provide
upper limits of $\approx$250~AU in diameter for MM1 and MM2.
Using Eq.~\ref{diskmass} in the optically thin limit gives estimates
of the disk masses for MM1 and MM2 of $0.09-0.24$~M$_{\odot}$ and
$0.12-0.33$~M$_{\odot}$, respectively, again assuming the
characteristic dust temperature to be $100-40$~K. Since the spectral
indices indicate that the emission is optically thick in both cases,
these masses should be treated as lower limits.

\begin{figure}
\centering{\includegraphics[width=8cm]{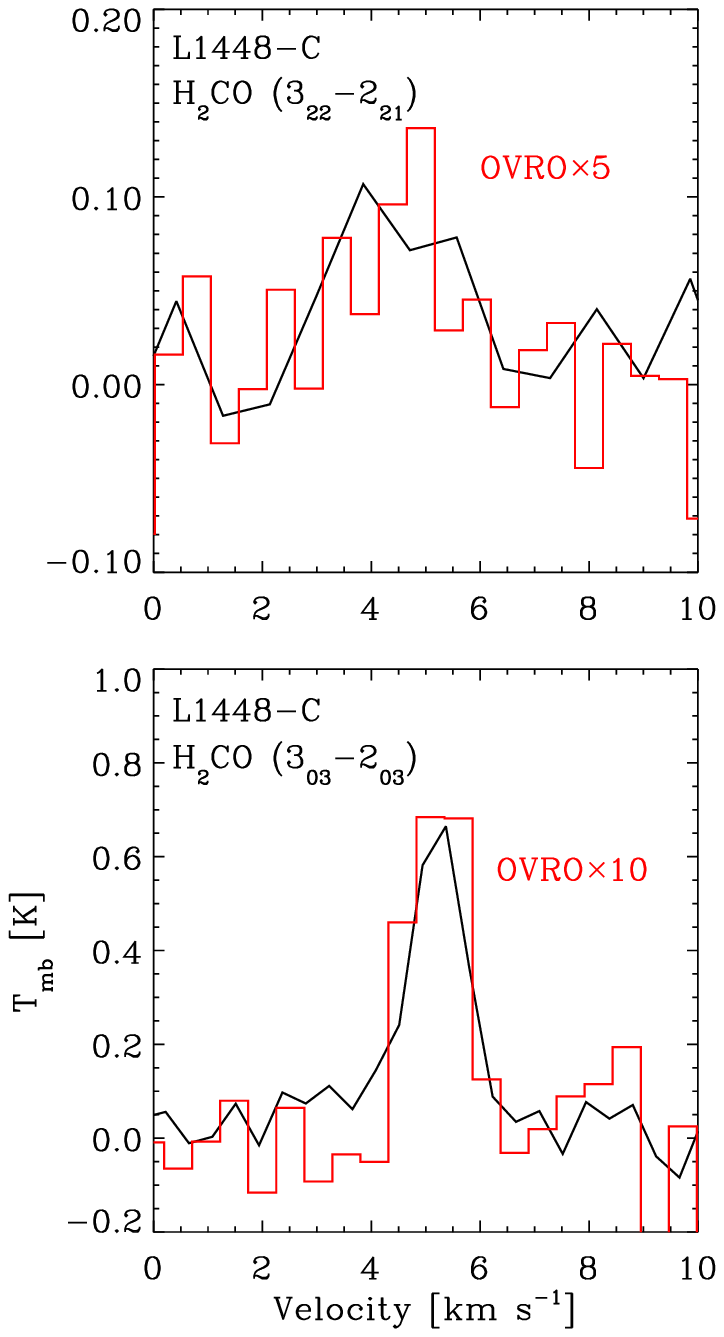}}
\caption{Comparison between the H$_2$CO line emission towards
\object{L1448--C}, at the source position, from JCMT single-dish
observations (line diagram) and OVRO interferometric observations
(histogram) restored with the JCMT beam (22\arcsec). The OVRO
spectrum has been scaled in order to account for the flux seen in
the JCMT spectrum.}
\label{compare_l1448}
\end{figure}

\section{H$_2$CO emission: morphology and abundance
structure}
\label{h2co}
\subsection{\object{L1448--C}}
\label{l1448_h2co_sec}
The maps of the H$_2$CO $3_{03}\rightarrow 2_{02}$ and
$3_{22}\rightarrow 2_{21}$ emission toward \object{L1448--C} are shown
in Fig.~\ref{l1448_h2co}, separated into blue ($4-5$\,km\,s$^{-1}$)
and red ($5-6$\,km\,s$^{-1}$) components. The $3_{03}\rightarrow 2_{02}$
emission appears to be slightly resolved with an extension to the
south along the direction of the outflow. The H$_2$CO
$3_{22}\rightarrow 2_{21}$ is detected only at the source
position. The velocity structure hints that the emission is related to
the known large scale outflow, although the velocities are
significantly lower than the high velocity (typically
$30-60$\,km\,s$^{-1}$) outflow seen in CO and SiO.

As for the continuum data, care has to be taken when interpreting the
line interferometer maps due to the low sensitivity to weak large
scale emission. A direct comparison between the single-dish spectrum
and that obtained from the interferometer observations restored with
the single-dish beam is shown in Fig.~\ref{compare_l1448}. The
interferometer picks up only $\sim10-20$\% of the single-dish flux,
suggesting that the extended cold material is resolved out by the
interferometer and that a hotter, more compact, component is
predominantly picked up. Within the considerable noise and coarsened
spectral resolution, the line profiles are consistent with those
obtained at the JCMT.

The $3_{22}\rightarrow 2_{21}$/$3_{03}\rightarrow 2_{02}$ line ratio
is sensitive to temperature (e.g., van Dishoeck et al.\ 1993\nocite{Dishoeck93}, Mangum
\& Wootten 1993\nocite{Mangum93}), especially in the regime of $50-200$\,K.  For
\object{L1448--C}, the interferometer data give a ratio of $0.68\pm
0.39$ indicating the presence of hot gas with $T\gtrsim 70$~K. For
comparison, the single-dish line ratio is $0.12\pm 0.04$ \citep{Maret04}, 
corresponding to $T\approx 20-30$~K.  Optical depth
effects and abundance variations with radius (i.e., temperature) can
affect this ratio, however, so that more detailed radiative transfer
modeling is needed for a proper interpretation.

Just as for the continuum data, the analysis of the H$_2$CO
interferometer data requires a detailed model of emission from the
more extended envelope as a starting point. Such a model has been
presented by \citet{Maret04} based on multi-line single-dish
observations.  Those data have been re-analyzed in this work using the
Monte Carlo radiative transfer method and molecular data adopted in
\citet{Schoeier02}. The density and temperature structures are taken
from \citet{Joergensen02} (see also \S\ref{l1448_dust}) assuming the
gas temperature to be coupled to that of the dust. The lines are
assumed to be broadened by turbulent motions in addition to
thermal line broadening. The adopted value of the turbulent velocity
is 0.7~km\,s$^{-1}$ \citep{Joergensen02}.  Considering only the
para-H$_2$CO data, a good fit ($\chi^2_{\mathrm{red}}=0.8$)
is obtained using a constant para-H$_2$CO abundance of $6\times 10^{-10}$
throughout the envelope.  A similarly good fit
($\chi^2_{\mathrm{red}}=1.0$) can be made to the ortho-H$_2$CO single-dish
data using an abundance of $9\times 10^{-10}$. The uncertainty of
these abundance estimates is approximately 20\% within the adopted
model.  The abundance of H$_2$CO  is $\sim$50\%
larger than that derived for the outer envelope of \object{IRAS
16293--2422} \citep{Schoeier02}.

As shown by \citet{Maret04}, a different interpretation is possible
within the same physical model if the ortho-to-para ratio is forced to
be equal to 3 and if a different velocity field is used.  If the gas
is assumed to be in free-fall toward the 0.5 M$_{\odot}$ protostar
with only thermal line broadening (i.e., no additional turbulent
velocity field), evidence of a huge abundance jump ($>$1000) can be
found for \object{L1448--C}.  The location of this jump is at the
100~K radius of the envelope, where thermal evaporation can take
place.  For L1448--C, this radius lies at 33 AU or 0$\farcs$15, and the
interferometer data can be used to test these two different
interpretations.

\begin{figure*}
\centering{\includegraphics[width=14cm]{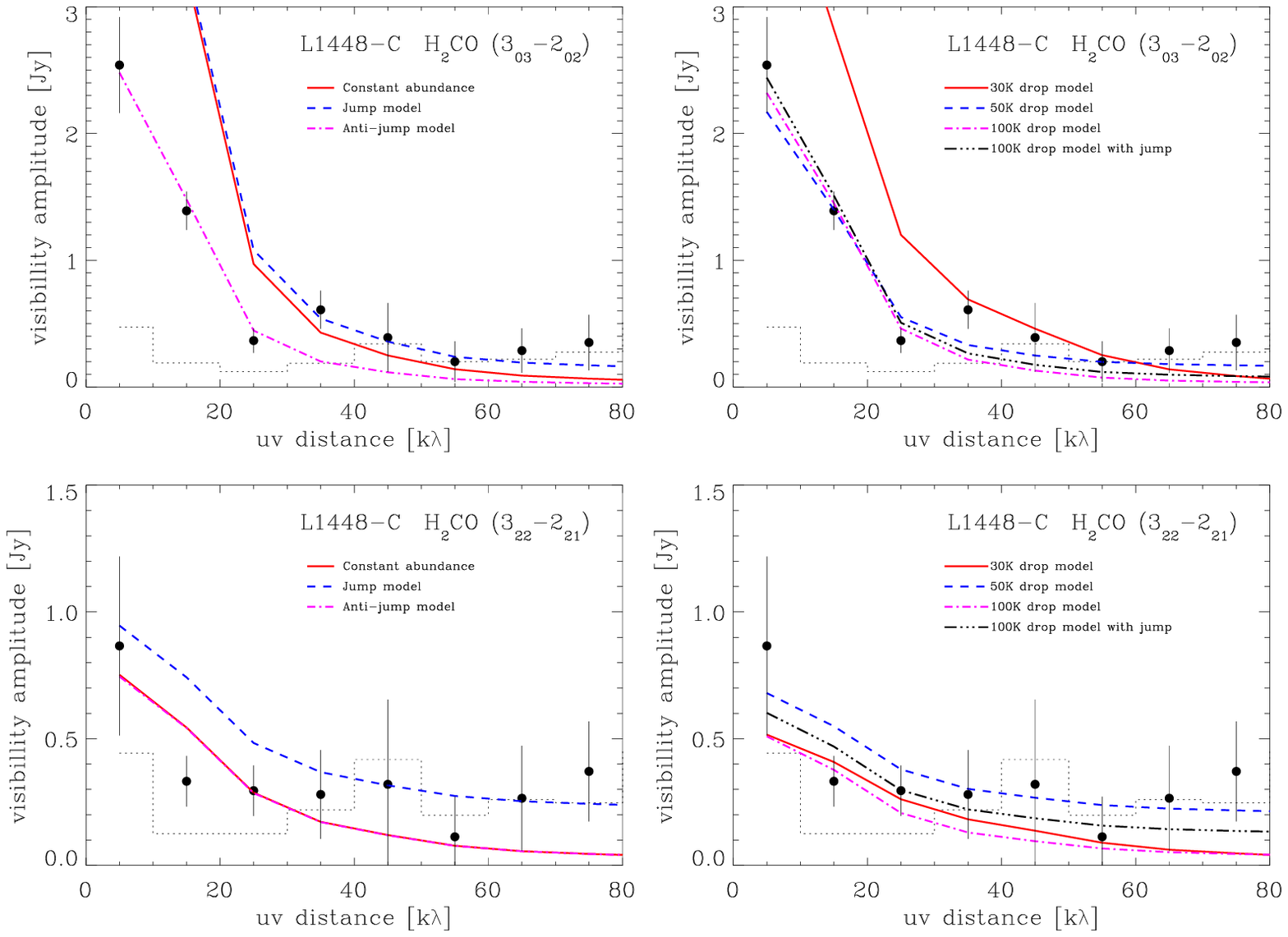}
\caption{Visibility amplitudes of the observed H$_2$CO line emission
obtained at OVRO toward \object{L1448--C} as functions of the
projected baseline length.  The observations, averaged from 4 to
6~km\,s$^{-1}$ and binned to 10~k$\lambda$, are plotted as filled
symbols with 1$\sigma$ error bars. The dotted histogram represents the 
zero-expectation level.
Also shown is the result of
applying the same ($u,v$) sampling to the envelope model for various scenarios 
for the H$_2$CO abundance distribution. See text for further details.
}
\label{l1448_h2co_vis}}
\end{figure*}

\begin{figure*}
\centering{\includegraphics[width=14cm]{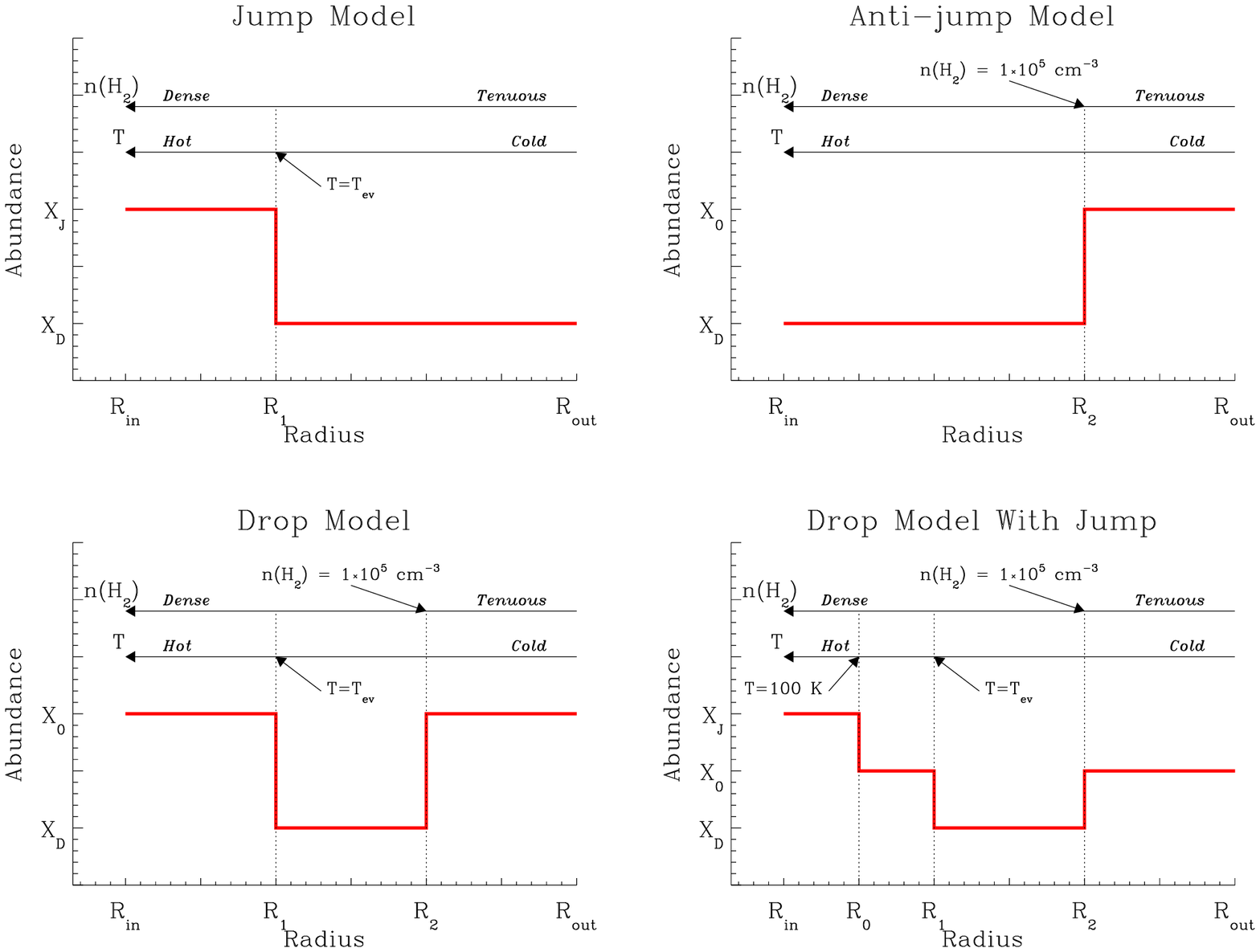}
\caption{Abundance profiles in various scenarios.}
\label{scenarios}}
\end{figure*}

Fig.~\ref{l1448_h2co_vis} shows the observed H$_2$CO visibility
amplitudes toward \object{L1448--C}. The emission has been averaged
over the full extent of the line ($4-6$~km\,s$^{-1}$). Although the
signal-to-noise is low, the emission is clearly resolved meaning that
hot H$_2$CO extends to scales larger than $1\arcsec$.
Fig.~\ref{l1448_h2co_vis} also presents the model predictions
assuming a constant para-H$_2$CO abundance
of $6\times 10^{-10}$, consistent with the
single-dish data (solid line). 
 The quality of the fit is measured using a
$\chi^2$ statistic for those visibility amplitudes that are above the
zero-expectation limit.  From the fit to the observed visibilities in
Fig.~\ref{l1448_h2co_vis} it is evident that such a constant H$_2$CO
abundance throughout the envelope cannot reproduce the interferometer
data  ($\chi^2_{\mathrm{red}}=26.8$): the $3_{03}-2_{02}$ emission is 
overproduced on
short baselines and underproduced on the longest baselines, while
the $3_{22}\rightarrow 2_{21}$/$3_{03}\rightarrow 2_{02}$ model
ratio is much lower than the observations.

One possible explanation is that the para-H$_2$CO abundance
drastically increases from $6\times10^{-10}$ in the outer cool parts
of the envelope to $7\times10^{-7}$ when $T>100$~K, in accordance with
the analysis of \citet{Maret04} and the results for
\object{IRAS~16293--2422} \citep{Ceccarelli00b,Schoeier02}. 
This situation is similar to that for the point source needed to
explain the continuum emission in \S\ref{dustmodelling}, since the
region where $T>100$~K is only about $0\farcs15$ (33~AU) in
radius. Enhancing the abundance in this region will therefore
correspond to adding an unresolved point source. The
visibility amplitudes for this model 
are presented in
Fig.~\ref{l1448_h2co_vis} (dashed line) and agree better with
observations on longer baselines than the constant abundance model,
 but the overall fit is actually slightly worse
($\chi^2_{\mathrm{red}}=31.4$).
Allowing for a jump at temperatures as low as 50~K cannot be ruled out
in the case of \object{IRAS 16293--2422} \citep{Schoeier02,
Doty04}. This would extend the region of warm material to scales of
$1\arcsec$ ($\sim 200$)~AU, but this still cannot explain the emission
on the shortest baselines, i.e.,  on scales of $\sim 10-20\arcsec$.

Similarly, the observed emission sampled by the longer baselines could
be caused, in part, by H$_2$CO emission coming from the warm
circumstellar disk that was suggested to explain the compact dust
emission in \S\ref{l1448_dust}.  Here we adopt a disk temperature of
100~K and a corresponding mass (Eq.~\ref{diskmass}) of
0.016~M$_{\odot}$. Further, assuming a diameter of 70~AU, similar to
the size of the region where the temperature is higher than 100~K and
consistent with the upper limit from observations, the spherically
averaged number density of H$_2$ molecules is
$5.3\times10^{9}$~cm$^{-3}$.  While the temperature is similar to that
found in the inner envelope, the density scale is $\sim 10-50$ times
higher. It is found that an H$_2$CO abundance of $4\times10^{-9}$ in
the disk can explain the observed visibilities on the longer
baselines.  The abundance in the disk is about 7 times larger than
found from single-dish modelling of the envelope alone, but more than
an order of magnitude lower than what the thermal evaporation model
predicts. The H$_2$CO abundance in the disk depends critically on the
assumed disk properties; e.g., the disk mass used above is only a
lower limit and the actual value may be an order of magnitude
higher. This would remove the need for a H$_2$CO abundance jump all
together. Still, since the disk is unresolved this does not alleviate
the problem of the relatively strong H$_2$CO $3_{22}\rightarrow
2_{21}$ line emission on the shortest baselines.

In their study of a larger sample of low-mass protostars,
\citet{Joergensen02} show that CO is significantly depleted in deeply
embedded objects such as \object{L1448--C}.  They also found that
intensities of low excitation $J=1\rightarrow 0$ lines are not
consistent with constant abundances derived on basis of the higher $J$
lines. Similar trends are seen for other molecular species such as
HCO$^+$ and HCN by \citet{Joergensen04c}, who suggest that this is
caused by the fact that the time scale for freeze-out of CO and other
species is longer than the protostellar lifetime in the outer
envelope. This leads to a `drop' abundance profile (see Fig.\
\ref{scenarios}), with CO frozen out in the cold region of the
envelope, but with standard or enhanced abundances in the outermost
low density cloud and in the inner warmer envelope.  The region over
which CO is frozen out is determined by the outer radius $R_2$ at
which the density is high enough that the freeze-out time scale is
short compared with the protostellar lifetime and the inner radius
$R_1$ at which the temperature is low enough that CO does not
immediately evaporate from the grain ice mantles. Photodesorption may
also play a role in the outermost region. In order for the freeze-out
time scale to be shorter than $\sim 10^4$~years, the density should be
higher than $\sim 10^5$~cm$^{-3}$.

The H$_2$CO abundance profile is expected to follow at least
partially that of CO, because destruction of gas-phase CO by
He$^+$ can be a significant source of atomic carbon and oxygen:
\begin{equation}
{\rm He}^{+} + {\rm CO} \rightarrow {\rm C}^+ + {\rm O} + {\rm He}
\end{equation}
For the typical densities and temperatures in the
outer region, H$_2$CO is mainly formed through the
reaction:
\begin{equation}
{\rm CH_3}+{\rm O} \rightarrow {\rm H_2CO}+{\rm H}
\end{equation}
so the H$_2$CO abundance should drop in regions where CO is frozen
out. Indeed, \citet{Maret04} found a clear correlation between the
H$_2$CO and CO abundances in the outer envelopes where both molecules
are depleted, for a sample of eight class~0 protostars.  Such an
effect would show up in the comparison of interferometer and
single-dish data: the interferometer data are mainly sensitive to the
$1-10\arcsec$ region of the envelope, where CO is frozen out. The
single-dish lines, however, either probe the outer regions of the
envelope (low excitation lines), where they quickly become optically
thick, or the inner regions (high excitation lines) which are
unaffected by the depletion.

In order to test this effect  several models in which H$_2$CO is
depleted over roughly the same region as CO were considered (see Fig.~\ref{scenarios}). 
Each
model was required to simultaneously reproduce all the available
multi-transition single-dish data to better than the 3$\sigma$ level.
First, an `anti-jump' model was considered 
where the abundance drops from an initial undepleted value, $X_0$, to
$X_{\mathrm D}$ when the H$_2$ density is larger than
10$^5$~cm$^{-3}$.  As can be seen in Fig.~\ref{l1448_h2co_vis} using
$X_0=5\times10^{-9}$ with a drop to $X_{\mathrm D}=6\times10^{-10}$
drastically improves the fit to the observed $3_{03}\rightarrow
2_{02}$ line emission due to opacity effects.  The optically thin
$3_{22}\rightarrow 2_{21}$ line emission is unaffected by this.  
The overall fit for this model is good, $\chi^2_{\mathrm{red}}=2.0$.

Next, a `drop' H$_2$CO abundance profile was introduced to simulate
the effects of thermal evaporation in the inner warm part of the
envelope.  First, $T_{\rm ev}$ was taken to be 30~K (see discussion in
\citet{Joergensen02}), roughly the evaporation temperature of CO.
 For $T_{\rm ev}=30$~K the jump is located at 
$R_1=7.5\times 10^{15}$~cm ($2\farcs3$). A para-H$_2$CO abundance 
of $X_0=5\times10^{-10}$ with a drop to $X_{\mathrm D}=3\times10^{-10}$
in the region of CO depletion provides the lowest $\chi^2$ and
is consistent with the single-dish data. However, the fit to
the interferometer data is not good, $\chi^2_{\mathrm{red}}=14.6$.
In particular the $3_{03}\rightarrow 2_{02}$ line emission comes out
too strong in the model since $X_0$ is not allowed to increase enough
(constrained by the single-dish data) to become optically thick as in
the anti-jump model. The `drop' model does, however, provide a good
description of the interferometer emission at both long and short baselines 
if the H$_2$CO abundance remains low out to $T\gtrsim 50$~K.
For $T_{\rm ev}=50$~K ($R_1=1.9\times 10^{15}$~cm; $0\farcs58$),
$X_0=5\times10^{-9}$ and $X_{\mathrm D}=4\times10^{-10}$ provides a
good fit with $\chi^2_{\mathrm{red}}=1.4$. Raising $T_{\rm ev}$ to
100~K ($R_1=5.0\times 10^{14}$~cm; $0\farcs15$)
provides a slightly worse fit, $\chi^2_{\mathrm{red}}=1.8$, for
$X_0=4\times10^{-9}$ and $X_{\mathrm D}=4\times10^{-10}$.  $X_0$ is
forced to be in the range $4-5\times10^{-9}$ from the single dish
data, so for $T_{\rm ev}=100$~K less flux is obtained at shorter
baselines than compared with the model where $T_{\rm ev}=50$~K because
of the smaller emitting volume.  Note that these models predict a
high $3_{22}\rightarrow 2_{21}$/$3_{03}\rightarrow 2_{02}$ line ratio
at large scales where the gas temperature is only $\sim$20~K, due to
the fact that the $3_{03}\rightarrow 2_{02}$ line in the outer
undepleted region becomes optically thick. On the longest baselines,
compact emission from either a disk or an abundance jump would still
be consistent with the observations.  For example, an additional
jump of a factor 10 when $T>100$~K ($R_0=R_1=5\times 10^{14}$~cm; 
$X_{\mathrm{D}}=4\times10^{-10}$;
$X_{\mathrm{0}}=4\times10^{-9}$; $X_{\mathrm{J}}=4\times10^{-8}$; see
Fig.~\ref{scenarios}) results in an additional $\approx 0.1$~Jy on all
baselines and improves the overal fit, $\chi^2_{\mathrm{red}}=1.4$,
for evaporation at this higher temperature (see
Fig.~\ref{l1448_h2co_vis}). However, since the observed signal is close to
the zero-expectation level, no strong conclusions on the presence of
this additional jump can be made.

\begin{figure}
\centering{\includegraphics[width=8cm]{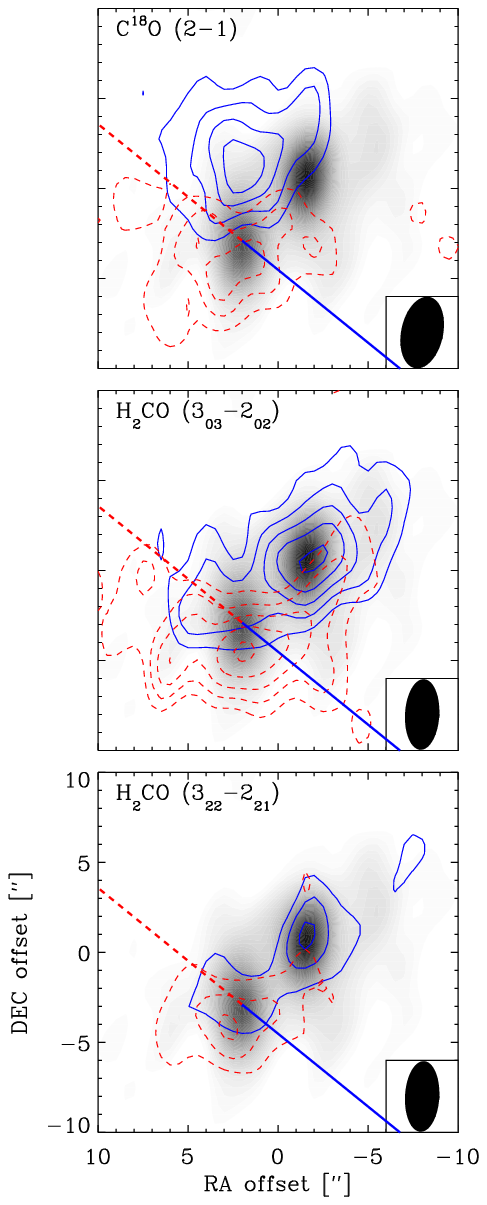}
\caption{OVRO interferometer maps of \object{IRAS 16293--2422} in
C$^{18}$O and H$_2$CO line emission (contours) overlayed on the
1.37\,mm continuum emission (greyscale).  The molecular line emission
has been separated into red (dashed lines) and blue (solid lines)
components (see text for details). Contours start at
1.8\,Jy\,beam$^{-1}$\,km\,s$^{-1}$ for C$^{18}$O and at
0.9\,Jy\,beam$^{-1}$\,km\,s$^{-1}$ for H$_2$CO.  The first contour
corresponds to the '2$\sigma$'-level as estimated from the maps and
each successive contour denotes an increase of 2$\sigma$. Also
indicated is the direction of the large scale CO outflow.}
\label{line_iras}}
\end{figure}

To summarize: while a constant abundance model can explain the H$_2$CO
emission traced by the single-dish data, it underproduces the emission
in the interferometer data on the longest baselines and produces too much
$3_{03}-2_{02}$ line emission on the shorter baselines. Adding a compact
source of emission either through a hot region of ice mantle
evaporation or a circumstellar disk  does not
provide a better overall fit.
A `drop' profile in which the H$_2$CO abundance largely follows that
of CO alleviates these problems, explaining at the same time the
emission seen by single-dish and the structure of the emission as
traced by the interferometer.  The relatively high H$_2$CO
$3_{22}-2_{21} /3_{03}-2_{02}$ ratio at scales of 1 to 10\arcsec\ is
in this scenario caused by a combination of high optical depth of the
$3_{03}-2_{02}$ line in the outermost region, and a low H$_2$CO
abundance in the cold dense part of the envelope where CO is frozen
out.  The best-fit abundances in each of these scenarios are
summarized in \S\ref{discussion} and compared with those obtained from
the similar analysis performed for \object{IRAS 16293--2422} in
\S\ref{iras_h2co}.

Finally, it should be noted that there may still be other explanations
for the high $3_{22}-2_{21} /3_{03}-2_{02}$ line ratio on short
baselines. If the H$_2$CO emission originates from low-velocity
entrained material in regions where the outflow interacts with the
envelope, the gas temperature may be increased due to weak
shocks. Alternatively, the gas temperature can be higher than that of
the dust due to heating by ultraviolet or X-ray photons from the
protostar which can escape through the biconical outflow cavity and
scatter back into the envelope at larger distances (cf.\ Spaans et al.
1995). Detailed quantitative modeling of these scenarios requires a
good physical model of the heating mechanisms and a 2-D radiative
transfer and model analysis, both of which are beyond the scope of
this paper. In either scenario, the general envelope emission
described above still has to be added, and may affect the line ratios
through opacity effects.

\begin{figure*}
\centering{\includegraphics[width=14cm]{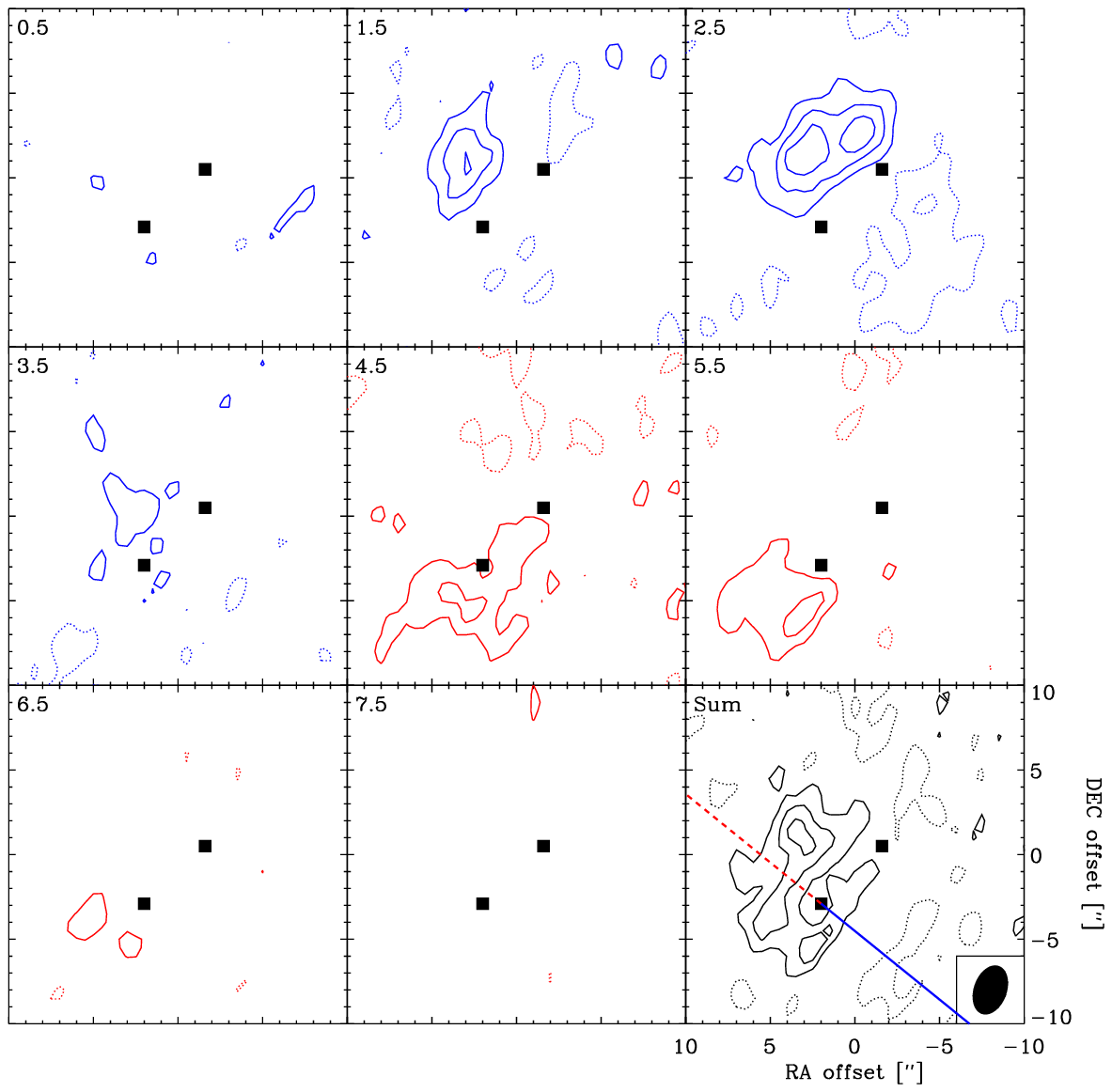}
\caption{OVRO interferometer maps of C$^{18}$O $J=2\rightarrow 1$ line  
emission
(contours) from \object{IRAS 16293--2422}. The C$^{18}$O emission has
been separated into bins of 1~km\,s$^{1}$ and contours are in steps
of 0.8~Jy\,beam$^{-1}$\,km\,s$^{-1}$ (2$\sigma$). Dotted contours
indicate negative values.  In the panel with the velocity-integrated
line intensity the contours start at
2.0~Jy\,beam$^{-1}$\,km\,s$^{-1}$ (2$\sigma$). Also indicated are the  
positions
of the two continuum sources as well as  the red- (dashed) and  
blue-shifted (solid) parts
of the large scale  outflow.}
\label{c18o}}
\end{figure*}

\subsection{IRAS 16293--2422}

\subsubsection{C$^{18}$O emission}
\label{c18o_iras}
The overall C$^{18}$O $J=2\rightarrow 1$ line emission for
\object{IRAS 16293--2422} obtained at OVRO is presented in
Fig.~\ref{line_iras}, with the channel maps shown in Fig.~\ref{c18o}.
The emission is clearly resolved and shows a $\sim$6$\arcsec$
separation between the red ($4-7$~km\,s$^{-1}$) and blue
($1-4$~km\,s$^{-1}$) emission peaks. The direction of the red-blue
asymmetry is roughly perpendicular to the large scale CO outflow
associated with MM1 (Walker et~al.\ 1988\nocite{Walker88}, Mizuno
et~al.\ 1990\nocite{Mizuno90}, Stark et~al.\ 2004\nocite{Stark04}),
and may be indicative of overall rotation of the circumbinary material
encompassing both MM1 and MM2. The morphology and velocity structure
is also consistent with a large ($\sim$1000~AU diameter) rotating
gaseous disk around just MM1, however. Such large rotating gaseous
disks have been inferred around other sources, including the class I
object L1489 \citep{Hogerheijde01} and the classical T Tauri star DM
Tau \citep{Dutrey97}, although their C$^{18}$O emission is usually too
weak to be detected.  The C$^{18}$O emission clearly avoids the region
between the binaries, consistent with the conclusion from the
continuum data that this region is void (see \S 3.2).

\begin{figure}
\centering{\includegraphics[width=8cm]{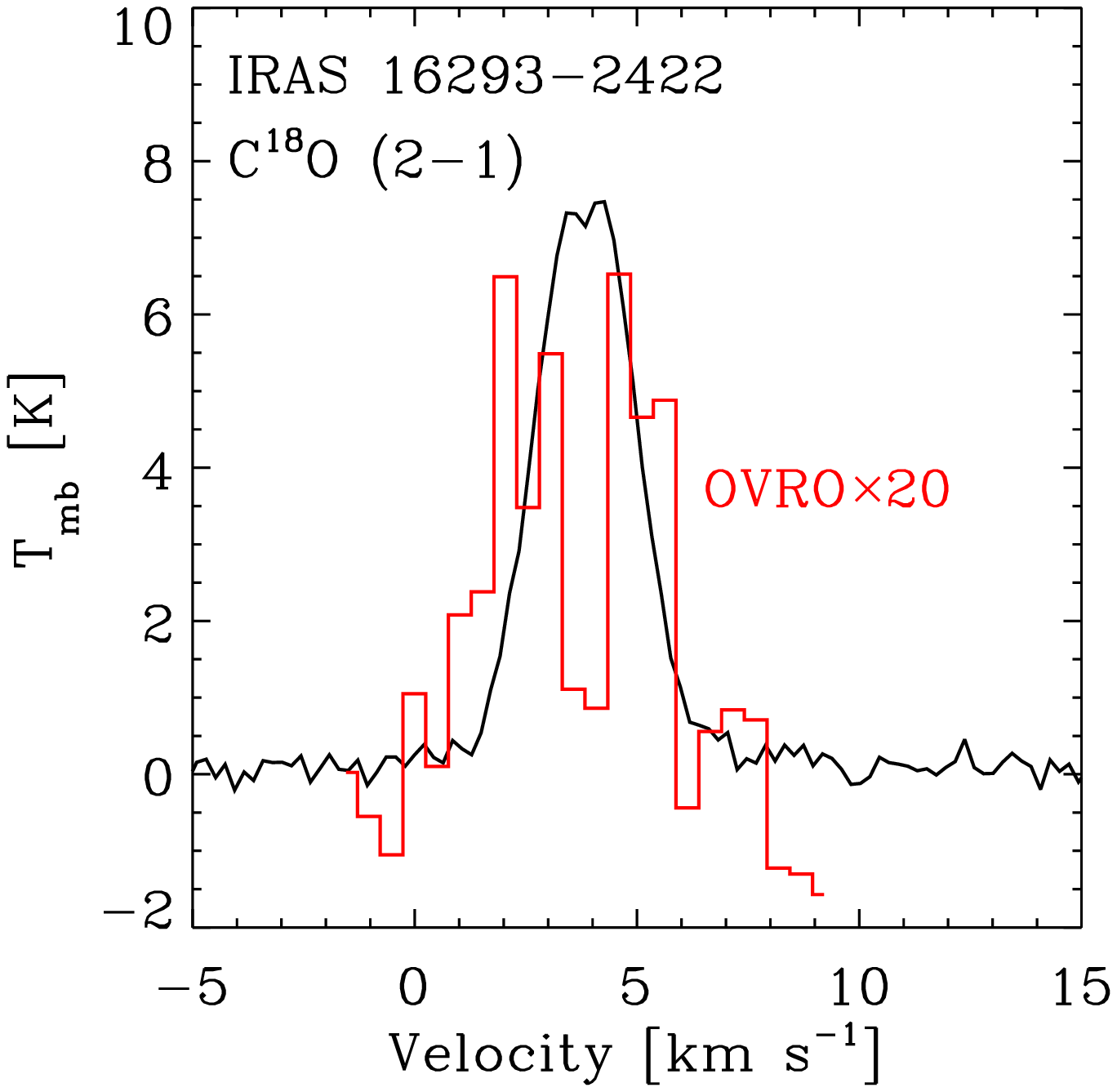}
\caption{Comparison between the C$^{18}$O $J=2\rightarrow 1$ line
emission towards \object{IRAS 16293--2422} at the source position,
from JCMT single-dish observations (line diagram) and OVRO
interferometric observations (histogram) restored with the JCMT
beam (22\arcsec). The OVRO spectrum has been scaled in order to
account for the flux seen in the JCMT spectrum.}
\label{c18o_compare}}
\end{figure}

In Fig.~\ref{c18o_compare} the interferometer data, restored with a
$22\arcsec$ beam, are compared with the JCMT single-dish flux. The
interferometer only recovers $\sim$5\% of the total single-dish flux,
mainly at extreme velocities. The interferometer is not sensitive to
the large scale static emission close to the cloud velocity of
$\approx$4~km\,s$^{-1}$.

The C$^{18}$O data can be used to further test the density structure
of the circumbinary envelope. However, as discussed in
\S\ref{l1448_h2co_sec} there is now growing evidence that CO is
depleted in the cool outer parts of the envelope so that extra care is
needed when deducing the density structure from CO observations alone.
\citet{Schoeier02} were able to reproduce single-dish CO observations
using a constant abundance of about $3 \pm 1\times 10^{-5}$ throughout
the envelope and with jump models
at 20~K, i.e., the characteristic evaporation
temperature, $T_{\mathrm{CO}}$, of pure CO ice. Recently,
\citet{Doty04} found that $T_{\mathrm{CO}}\approx 20$~K provided the
best fit in their chemical modelling of a large number of molecular
species.  Here the C$^{18}$O $J=2\rightarrow 1$ data are analyzed
assuming that the emission originates from 1) a `standard' envelope
centered on MM1; and 2) a circumbinary envelope with a cavity
centered between the positions of the protostars MM1 and MM2.

\begin{figure}
\centering{\includegraphics[width=8cm]{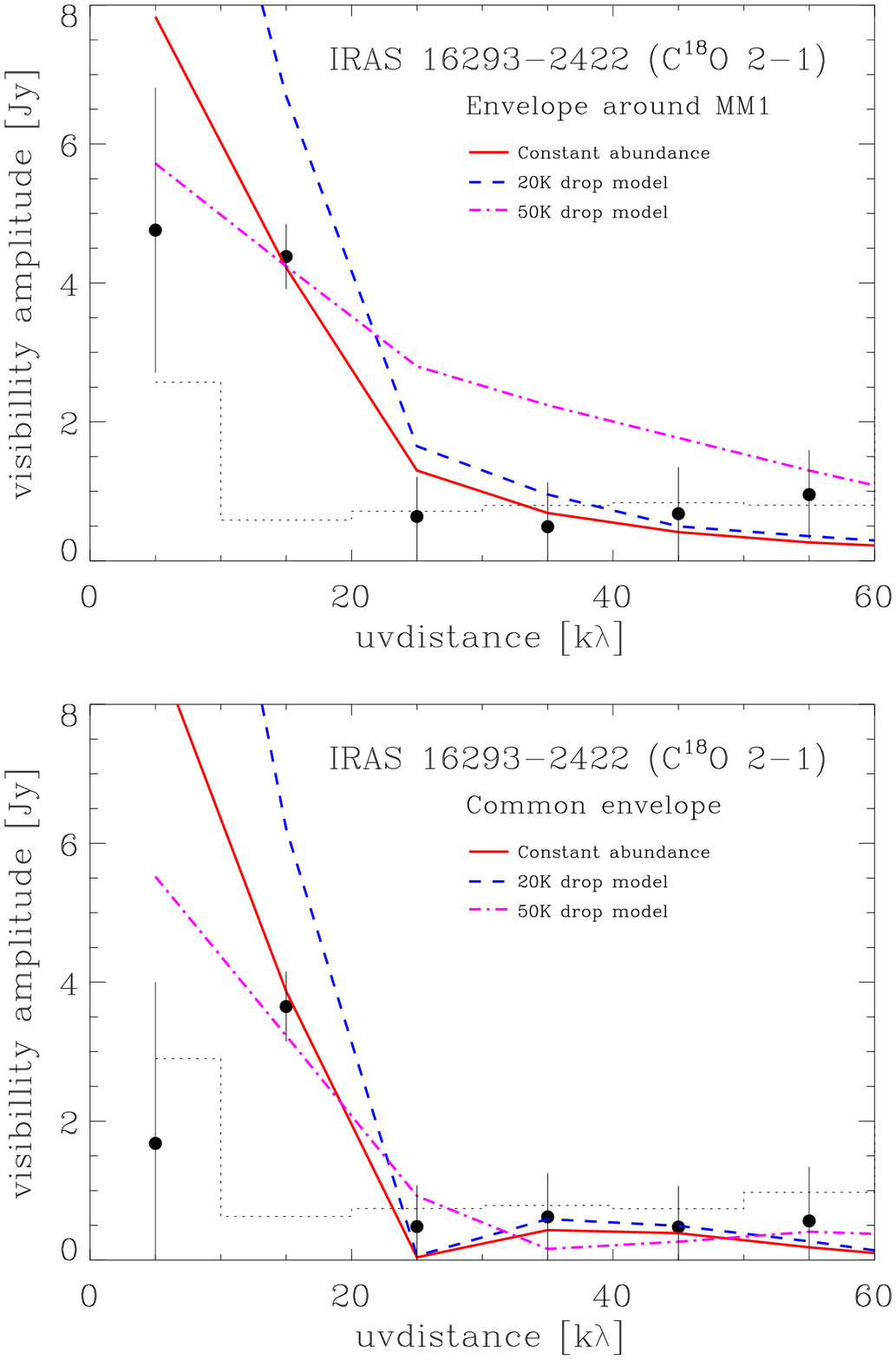}
\caption{Visibility amplitudes of the observed C$^{18}$O
$2\rightarrow1$ line emission obtained at OVRO toward \object{IRAS
16293--2422} as functions of the projected baseline length from the
phase center, taken to be at ($2\arcsec,-3\arcsec$) for the envelope
around MM1 (top) and ($0\farcs2,-1\farcs2$) for the common envelope
(bottom).  The observations, averaged over 1 to 7~km\,s$^{-1}$ and
binned to 10~k$\lambda$, are plotted as filled symbols with 1$\sigma$
error bars. The dotted histogram represents the zero-expectation
level.  Also shown is the result of applying the same ($u,v$) sampling
to the circumstellar model using a constant C$^{18}$O abundance of
$6\times 10^{-8}$ (solid lines) for an envelope centered on the
protostar MM1 and a common envelope with a cavity. Envelope models
where CO is depleted in a region where $n_{\mathrm{H_2}}>1\times
10^{5}$~cm$^{-3}$ and $T<T_{\mathrm{ev}}$ are also indicated as dashed
($T_{\mathrm{ev}}=20$~K; $X_0=8\times 10^{-8}$;
$X_{\mathrm{D}}=2\times 10^{-8}$) and dot-dashed
($T_{\mathrm{ev}}=50$~K; $X_0=2\times 10^{-7}$;
$X_{\mathrm{D}}=2\times 10^{-8}$) lines.}
\label{iras_c18o_vis}}
\end{figure}

Fig.~\ref{iras_c18o_vis} shows the result of applying the ($u,v$)
sampling of the observations to the C$^{18}$O envelope models. In both
the model centered on MM1 and the common envelope model (with a
cavity) the observed visibilities are relatively well reproduced using
the same constant C$^{18}$O abundance of $6\times10^{-8}$ (solid
lines) as obtained from the single-dish analysis, except perhaps at
the longest baselines.  Note that the density and temperature
structures are slightly different for the model with a cavity (see
\S\ref{dust_iras}).  Next, thermal evaporation models with a drastic
jump at $T_{\mathrm{CO}} = 20$~K, as suggested by \citet{Doty04}, were
considered. The abundance in the region of depletion, i.e., when $T <
20$~K and $n_{\mathrm{H_2}} > 1\times 10^{5}$~cm$^{-3}$, was fixed to
$2\times10^{-8}$, the upper limit obtained by \citet{Doty04}. An
undepleted C$^{18}$O abundance, $X_0$, of $8\times10^{-8}$
(corresponding to a total CO abundance of about $5\times10^{-5}$) is
found to provide reasonable fits to the observed visibilities in
Fig.~\ref{iras_c18o_vis} (dashed lines) as well as CO single-dish
data, and is also just consistent with the chemical modelling
performed by \citet{Doty04}.  This CO abundance is a factor of 2 to 5
lower than typical undepleted abundances. Assuming a still higher
evaporation temperature of 50~K brings $X_0$ for C$^{18}$O up to
$2\times10^{-7}$ (CO up to $1.1\times 10^{-4}$), more in line with
what is typically observed in dark clouds. \citet{Doty04} could not
rule out such high evaporation temperatures in their chemical modelling.
In all, the success of the envelope model in explaining both the
continuum emission as well as the observed CO emission is encouraging
and will aid the interpretation of the H$_2$CO observations.

The flux detected on the longer baselines provides a limit on the
maximum CO abundance in the disk. For a disk around MM1 with diameter
250~AU, temperature 100~K and the mass 0.09~M$_{\odot}$ (see
\S\ref{dust_iras}) the spherically averaged H$_2$ number density is
$6.4\times 10^{8}$~cm$^{-3}$. It is found that a C$^{18}$O abundance
of $\approx 2\times10^{-7}$, corresponding to a total CO abundance of about
$1.1\times 10^{-4}$, can account for the emission on the longest
baselines. Assuming a lower temperature in the disk of 40~K and the
correspondingly higher mass of 0.24~M$_{\odot}$ gives almost the same
upper limit to the amount of CO in the disk.

\begin{figure}
\centering{\includegraphics[width=8cm]{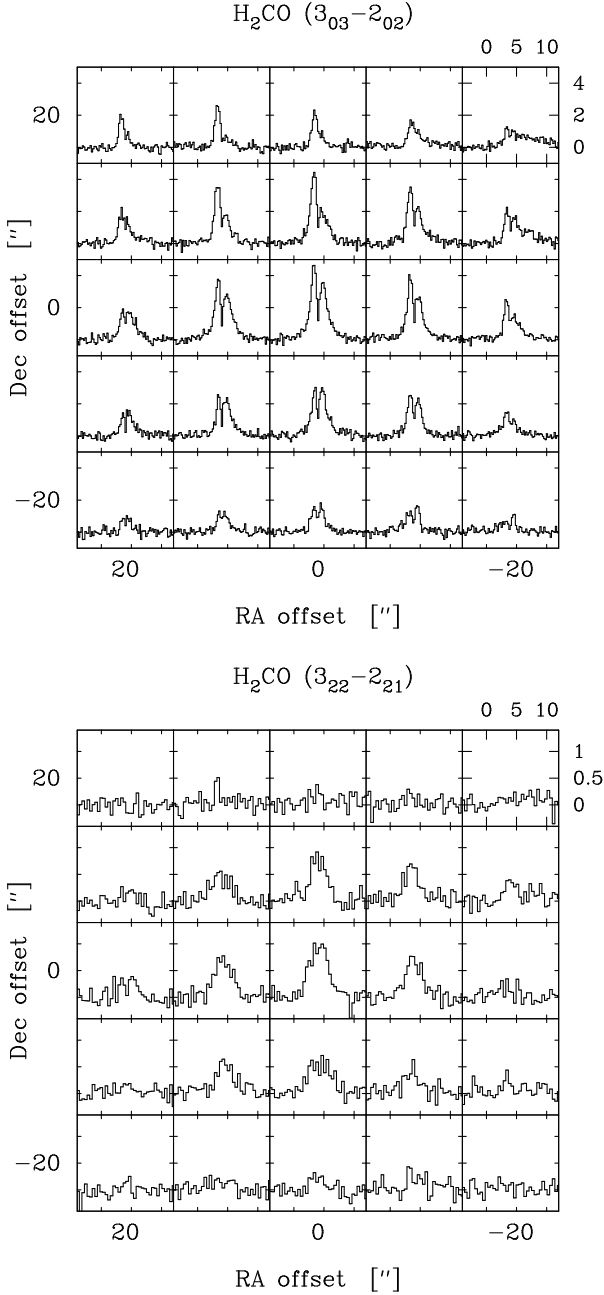}}
\caption{JCMT single-dish spectral maps of the H$_2$CO emission toward
\object{IRAS 16293--2422}. The velocity scale is in km\,s$^{-1}$ and
the intensity is the main beam brightness temperature in K as
indicated in the upper right subplot. The spectral resolution is
0.22~km\,s$^{-1}$. The $3_{22}\rightarrow 2_{21}$ line data have been
smoothed to a two times lower spectral resolution.}
\label{jcmt}
\end{figure}

\subsubsection{H$_2$CO emission}
\label{iras_h2co}
The single-dish observations (see Fig.~\ref{jcmt}) 
show that the H$_2$CO line emission is extended to scales of
$\sim$30$\arcsec$.  The single-dish $3_{22}\rightarrow
2_{21}$/$3_{03}\rightarrow 2_{02}$ line ratio is $\approx$0.2, suggesting
that a cold ($30-40$~K) envelope component dominates the single-dish
flux. In contrast, the interferometer $3_{22}\rightarrow
2_{21}$/$3_{03}\rightarrow 2_{02}$ line ratio is $0.69\pm 0.23$ for
the red-shifted emission near MM1 and $0.75\pm 0.32$ for the
blue-shifted emission close to MM2. This indicates that the
temperature is in excess of $\sim 150$~K assuming the density to be
at least $10^{6}$~cm$^{-3}$.  However, as noted for \object{L1448--C}, 
optical depth effects and abundance gradients can affect this ratio.

\begin{figure*}
\centering{\includegraphics[width=14cm]{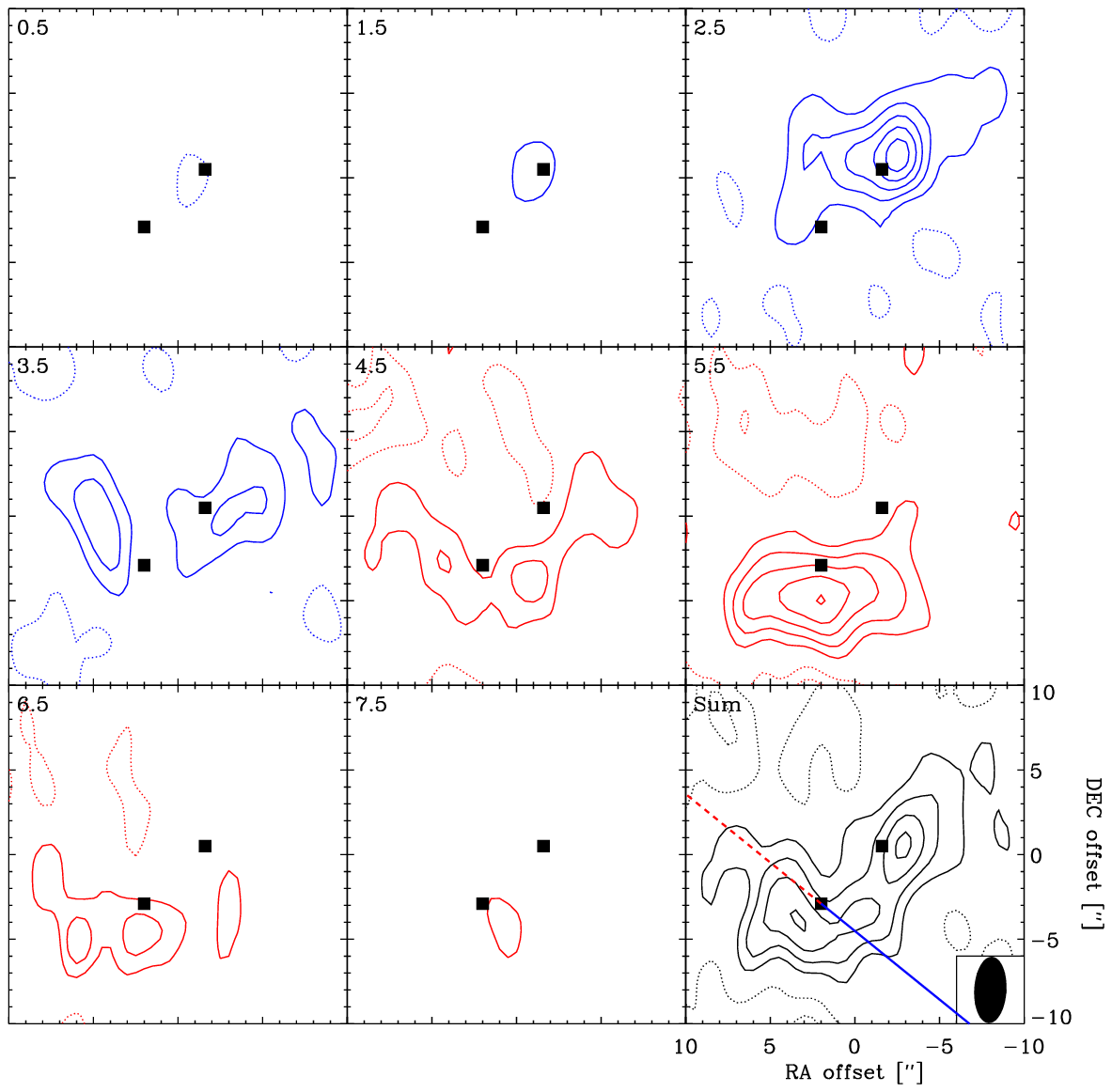}
\caption{OVRO interferometer maps of H$_2$CO $3_{03}\rightarrow
2_{02}$ line emission (contours) for \object{IRAS 16293--2422}. The
H$_2$CO emission has been separated into bins of 1\,km\,s$^{1}$ and
contours are in steps of 0.4\,Jy\,beam$^{-1}$\,km\,s$^{-1}$
(2$\sigma$). In the panel with the velocity-integrated line intensity
the contours start at 1.0\,Jy\,beam$^{-1}$\,km\,s$^{-1}$
(2$\sigma$). Also indicated are the positions of the two continuum
sources as well as the red- (dashed) and blue-shifted (solid) parts of
the outflow.}
\label{h2co_iras}}
\end{figure*}

The velocity channel maps of the H$_2$CO $3_{03}\rightarrow 2_{02}$
interferometer line emission obtained at OVRO are shown in
Fig.~\ref{h2co_iras} for \object{IRAS 16293--2422}, whereas the total
H$_2$CO $3_{22}\rightarrow 2_{21}$ and $3_{03}\rightarrow 2_{02}$ maps
are included in Fig.~\ref{line_iras}.  The emission is clearly
resolved and shows a $\sim$6$\arcsec$ separation between the red
($4-7$~km\,s$^{-1}$) and blue ($1-4$~km\,s$^{-1}$) emission
peaks. The direction of the red-blue asymmetry is again roughly
perpendicular to the large scale CO outflow associated with MM1, but
in contrast with the C$^{18}$O OVRO maps, no blue-shifted emission
close to MM1 is observed. Instead, the strongest blue-shifted emission
occurs close ($<$1$\arcsec$) to MM2. The red-shifted H$_2$CO emission
is found to the south of MM1, but again peaks closer to MM1 than the
C$^{18}$O emission does.

\begin{figure}
\centering{\includegraphics[width=7cm]{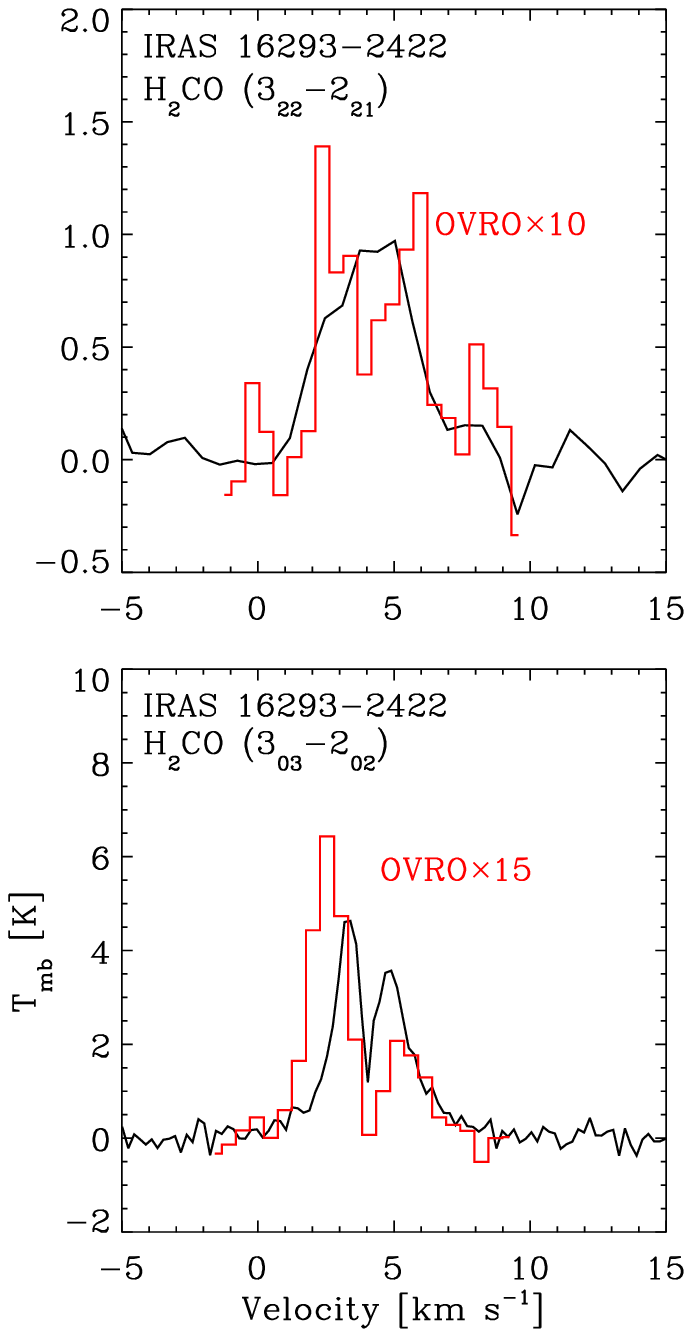}
\caption{Comparison between the H$_2$CO  line
emission towards \object{IRAS 16293--2422} at the source position,
from JCMT single-dish observations (line diagram) and OVRO
interferometric observations (histogram) restored with the JCMT
beam (22\arcsec). The OVRO spectrum has been scaled in order to
account for  the flux seen in the JCMT spectrum.}
\label{compare}}
\end{figure}

As for C$^{18}$O, only a small fraction ($\sim 5-20$\%) of the
single-dish flux is recovered by the interferometer (see
Fig.~\ref{compare}). The shapes of the H$_2$CO $3_{03}\rightarrow
2_{02}$ and $3_{22}\rightarrow 2_{21}$ lines are very similar to that
of C$^{18}$O $J=2\rightarrow 1$ (see Fig.~\ref{c18o_compare}), when
the interferometer data are restored with the JCMT beam.

\begin{figure*}
\centering{\includegraphics[width=14cm]{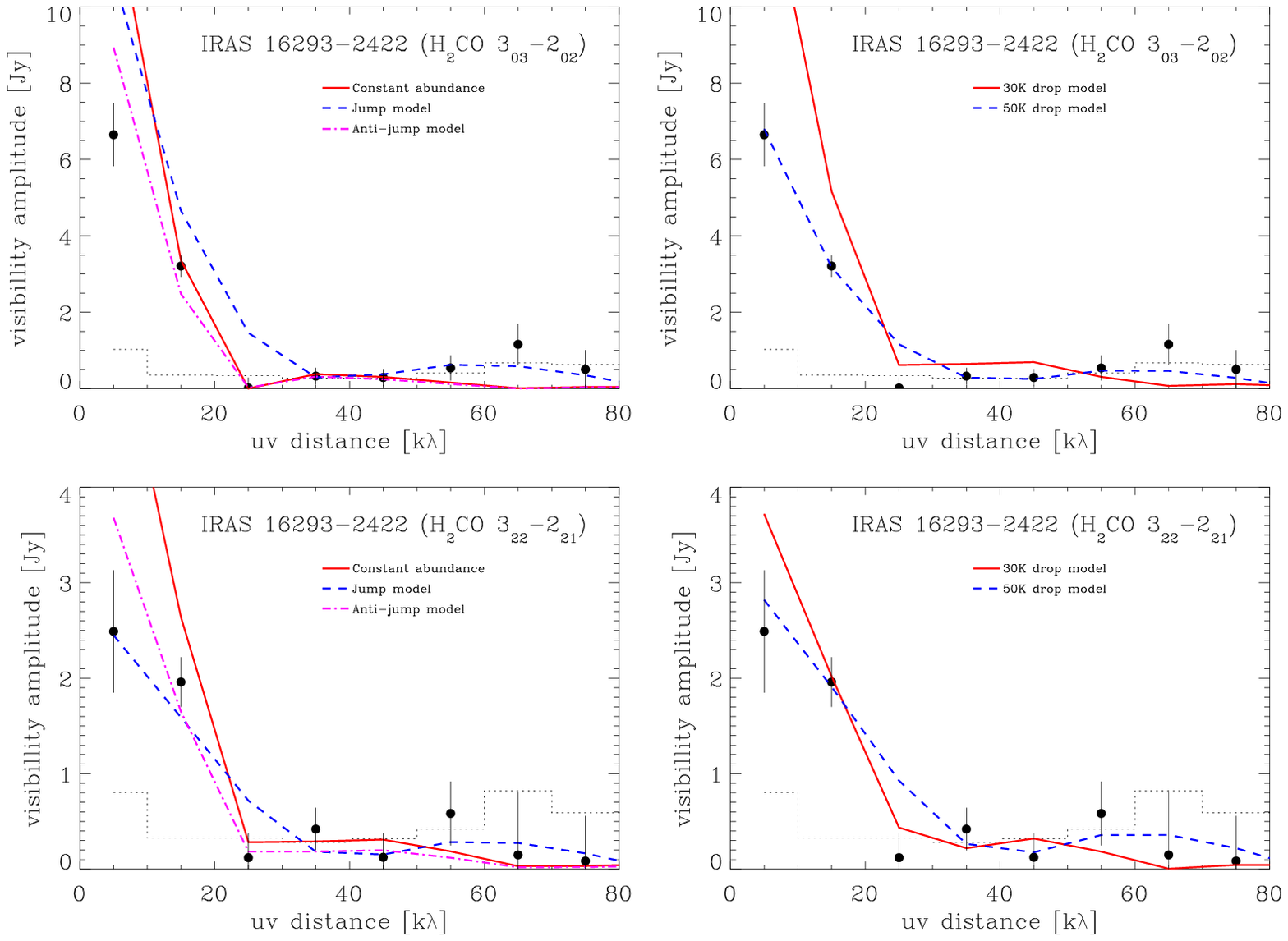}}
\caption{Visibility amplitudes of the observed H$_2$CO line emission
obtained at OVRO toward \object{IRAS 16293--2422} as functions of the
projected baseline length from the phase center, taken to be at
($0\farcs2,-1\farcs2$) for the common envelope.  The observations,
averaged over 1 to 7~km\,s$^{-1}$ and binned to 10~k$\lambda$, are
plotted as filled symbols with 1$\sigma$ error bars. The dotted
histogram represents the zero-expectation level.  Also shown are the
results of applying the same ($u,v$) sampling to the circumbinary
envelope model using various H$_2$CO abundance distributions. 
 All models are consistent with available multi-transition
single-dish data to within 3$\sigma$. See text 
and Fig.~\ref{scenarios} for
details on these models.}
\label{iras_h2co_uv}
\end{figure*}

The H$_2$CO emission is interpreted in terms of a common envelope
scenario using the cavity model that was seen to best reproduce the
emission in \S\ref{dust_iras}. Alternative models with the envelope
centered on MM1 or MM2 have been considered as well, but lead to the
same overall conclusions. H$_2$CO models with and without any
abundance jumps or drops are produced for this physical model, 
 following the
analysis performed for \object{L1448--C} (see
Fig.~\ref{scenarios}).  Applying the
($u,v$) sampling from the observations to these models produces
visibility amplitudes that are compared with observations in
Fig.~\ref{iras_h2co_uv}. 

Using a constant para-H$_2$CO abundance of $5\times10^{-10}$ derived
from the single-dish modelling performed in \citet{Schoeier02}
produces too much flux on the shorter baselines for both the
$3_{03}\rightarrow 2_{02}$ and $3_{22}\rightarrow 2_{21}$ transitions.
For \object{IRAS 16293--2422}, an abundance jump in the inner hot core
due to evaporation of ice mantles is well established
from multi-transition single-dish modelling \citep{Dishoeck95,
Ceccarelli00b,Schoeier02}. 
\citet{Schoeier02} constrain the location of the jump to
$\gtrsim 40$~K, with a jump in
abundance of one to three orders of magnitude depending on the
adopted evaporation temperature. 
It is found that a jump model with
$T_{\mathrm{ev}}=50$~K (at $R_1=6.2\times 10^{15}$~cm =  410~AU = $2\farcs6$) 
improves the fit to the data
($\chi^2_{\mathrm{red}}=7.1$) using $X_0=1\times10^{-10}$ and
$X_{\mathrm J}=3\times10^{-9}$ (Fig.~\ref{iras_h2co_uv}, left
panels). However, this model still produces too much
$3_{03}\rightarrow 2_{02}$ line emission on the shortest
baselines. This is similar to \object{L1448--C} where a better fit was
obtained by letting the abundance increase again in the outer region when
$n_{\mathrm{H_2}}<1\times 10^5$~cm$^{-3}$ to provide significant
optical depth in the $3_{03} \rightarrow 2_{02}$ transition.
Such an anti-jump model
with $X_{\mathrm D}=1\times10^{-10}$ and $X_0=3\times10^{-9}$ also
improves the fit ($\chi^2_{\mathrm{red}}=3.6$) in the case of
\object{IRAS 16293--2422}.

In the drop models two different evaporation temperatures are
considered: $T_{\mathrm{ev}} =50$~K  and
30~K (at $R_1=1.1\times 10^{16}$~cm =  740~AU = $4\farcs6$). As shown in Fig.~\ref{iras_h2co_uv}
(right panels) the 30~K model does not provide a good fit to the data
($\chi^2_{\mathrm{red}}=17.5$). Using $T_{\mathrm{ev}}=50$~K instead
allows $X_0$ to be higher and a near-perfect fit
($\chi^2_{\mathrm{red}}=0.4$) can be found to both the
$3_{03}\rightarrow 2_{02}$ and $3_{22}\rightarrow 2_{21}$ line
emission. In this case $X_{\mathrm D}=1\times10^{-10}$ and
$X_0=4\times10^{-9}$, i.e., the jump is of a factor 40.  A similar jump at 50 K was found by
\citet{Ceccarelli01} from their analysis of the H$2$CO single-dish data.
The best-fit abundances in each of these scenarios are
summarized in \S\ref{discussion} and compared with those obtained from
the similar analysis performed for \object{L1448--C} in
\S\ref{l1448_h2co_sec}.


The observed flux at the longest baselines gives an upper limit to the
para-H$_2$CO abundance in a disk around MM1. Using the
properties of the disk as in \S\ref{c18o_iras} an abundance of
$1\times10^{-9}$ is found to produce about 0.5~Jy on all baselines. 
 For the $\approx$30\% more massive disk around MM2 a slightly lower value 
for the para-H$_2$CO abundance is found.

Given the complexity of the H$_2$CO emitting region on small scales
($\lesssim$10$\arcsec$) the envelope model presented in
\citet{Schoeier02} is not adequate to fully describe the morphology of
the emission observed by the interferometer.  Here we simply note that
the envelope model can explain the observed flux if jumps in abundance
are introduced.  The fact that both the H$_2$CO and C$^{18}$O
observations can be well reproduced by the common envelope model with
a cavity further supports the finding from the continuum observations
that little material seems to be located at inter-binary scales.

\begin{table*}
\caption[]{Best fit H$_2$CO envelope models$^a$.}
  \label{models}
\begin{tabular}{lccccccccccc}
\hline\hline
\noalign{\smallskip}
\multicolumn{1}{c}{Model} & &
\multicolumn{4}{c}{\object{L1448--C}} & &
\multicolumn{4}{c}{\object{IRAS 16293--2422}}\\
  &&
 $X_{\mathrm D}$ &  $X_0$ & $X_{\mathrm J}$ & $\chi^2_{\mathrm{red}}$$^b$ &&
 $X_{\mathrm D}$ &  $X_0$ & $X_{\mathrm J}$ & $\chi^2_{\mathrm{red}}$$^b$ \\
\hline
\noalign{\smallskip}
Constant abundance   && $6\times 10^{-10}$ & --- & --- & 26.8 && $5\times 10^{-10}$ & --- & --- & 11.0 \\
Jump model$^c$          && $6\times 10^{-10}$ & --- & $6\times 10^{-07}$ & 31.4 && $1\times 10^{-10}$ & --- & $3\times 10^{-09}$ &  \phantom{0}7.1 \\
Anti-jump model            && $6\times 10^{-10}$ &  $5\times 10^{-09}$ & --- & \phantom{0}2.0 && $3\times 10^{-10}$ & $4\times 10^{-09}$ & --- &  \phantom{0}3.6\\
30~K drop model          && $3\times 10^{-10}$ &  $5\times 10^{-10}$ & --- & 14.6 &&$2\times 10^{-10}$ & $4\times 10^{-10}$ & --- &  17.5\\
50~K drop model          &&$4\times 10^{-10}$ &  $5\times 10^{-09}$ & --- & \phantom{0}1.4 &&$1\times 10^{-10}$ & $4\times 10^{-09}$ & --- &  \phantom{0}0.4 \\
100~K drop model        &&$4\times 10^{-10}$ &  $4\times 10^{-09}$ & --- & \phantom{0}1.8 && --- & --- & --- & ---\\
100~K drop model with jump       &&$4\times 10^{-10}$ &  $4\times 10^{-09}$ & $4\times 10^{-08}$ & \phantom{0}1.4 && --- & --- & --- & ---\\
\hline
\noalign{\smallskip}
\end{tabular}

\noindent
$^a$ All abundances refer to para-H$_2$CO only. \\
$^b$ Reduced $\chi^2$ from interferometer data. All models are consistent with available single-dish data to within the 3$\sigma$ level.\\
\noindent
$^c$ Jump at 100~K for \object{L1448--C} and 50~K for \object{IRAS 16293--2422}.\\

\end{table*}

\section{Origin of the H$_2$CO emission}
\label{discussion}
Here the results from the previous sections are summarized.
Various competing scenarios on the origin of the observed H$_2$CO
emission are discussed and predictions for future generation telescopes
are presented, illustrating their potential to distinguish competing
scenarios.

\subsection{Envelope and/or outflow emission?}
In \S\ref{h2co} the physical envelope models derived from single-dish
observations, and tested against interferometric continuum
observations in \S\ref{dustmodelling}, have used as a basis for
interpreting the observed H$_2$CO emission. The results are summarized
in Table~\ref{models}.  It is found that for both \object{IRAS
16293--2422} and \object{L1448--C}, the best fit to the H$_2$CO
interferometric observations is obtained with a `drop' abundance
profile, in which the H$_2$CO abundance is lower by more than an order
of magnitude in the cold dense zone of the envelope but is high in the
inner- and outermost regions. The outer radius of this `drop'-zone is
set by the distance at which the density drops below $10^5$ cm$^{-3}$;
the inner radius by the distance at which the temperature is above the
evaporation temperature $T_{\rm ev}$.  Indeed, such a `drop' model
with $T_{\rm ev}$=50~K gives very good $\chi^2$ fits and reproduces
both the single-dish and interferometer data. The fact that two lines
of the same molecule with different optical depths were observed
simultaneously with OVRO was essential to reach this conclusion.  The
presence of abundance enhancements in regions where $T\gtrsim 50$~K is
consistent with the detailed analysis of multi-transition single-dish
H$_2$CO observations, although the actual values derived here are
somewhat different. Interestingly, the inferred abundances $X_D$ and
$X_0$ for the best-fit 50~K drop models are comparable for the two
sources. The presence of additional jumps with $X_J>X_0$ in the
innermost hot core region or disk cannot be established with
the current data, but requires interferometer observations of higher
excitation lines (see \S\ref{predictions_sec}).

Can some of the enhanced H$_2$CO originate in the outflow?  The
red-blue asymmetry observed for \object{L1448--C} is consistent with
the high velocity outflow seen in CO and SiO, but the velocities
seen for H$_2$CO are significantly lower. One explanation is that the
emission originates in the acceleration region of the outflow,
estimated to be within $2\arcsec$ radius from the star
\citep{Guilloteau92}. An alternative, more plausible scenario is that
if H$_2$CO is associated with the outflow, but located in
low-velocity entrained material in regions where the outflow interacts
with the envelope, since the red-shifted emission appears to be
extended to scales much larger than 2$''$. This would be similar to
the case of HCO$^+$ \citep{Guilloteau92}.

For \object{IRAS 16293--2422} the morphology indicates emission in a
rotating envelope perpendicular to the direction of the (large-scale)
outflow.  However, the complicated physical structure of this
protobinary object on small spatial scales is not well represented by
our spherically symmetric model, so it is not possible to rule out a
scenario where some of the emission originates on larger scales due to
interaction with the outflow.

To quantify the role of outflows in producing H$_2$CO abundance
enhancements and liberating ice mantles, H$_2$CO interferometer data
at $\sim$1$''$ resolution for a larger sample of sources are needed to
investigate whether the velocity pattern is systematically oriented
along the outflow axis as in L1448--C. Also, higher sensitivity could
reveal whether the profiles have more extended line wings.

\subsection{Photon heating of the envelope?}

The current models assume that the gas temperature equals the dust
temperature. Detailed models of the heating and cooling balance of the
gas have indicated that this is generally a good assumption within a
spherically symmetric model \citep{Ceccarelli96, Doty97}.  However, if
the gas temperature were higher than the dust temperature in certain
regions, this would be an alternative explanation for the increased
$3_{22}-2_{21}$/$3_{03}-2_{02}$ ratio in the interferometer data on
short baselines. One possibility discussed in \S\ref{l1448_h2co_sec}
would be gas heating by ultraviolet (UV)- or X-ray photons which
impact the inner envelope and can escape through the biconical cavity
excavated by the outflow. If such photons scatter back into the
envelope, they can raise the gas temperature to values significantly
in excess of the dust temperature in part of the outer envelope
\citep[e.g.,][]{Spaans95}.  For \object{IRAS 16293--2422}, such
photons can further escape through the circumbinary cavity, so that
the photons from e.g., MM2 can influence the inner envelope rim around
MM1.  Since this model would affect the excitation of all molecules
present in this gas, not only H$_2$CO, it can be tested with future
multi-line interferometer data of other species. Moreover, the
presence of UV photons would have chemical consequences producing
enhanced abundances of species like CN in the UV-affected regions,
which should be observable. Note that in this scenario, the general
colder envelope still has to be added, which, as noted above, can
affect the line ratios.

\subsection{Disk emission?}
The detailed modelling of the continuum emission performed in
\S\ref{dustmodelling} reveals that there is compact emission in both
\object{IRAS 16293--2422} and \object{L1448--C} that cannot be
explained by the envelope model. The most likely interpretation is
that of accretion disks. For \object{L1448--C}, where the inner region
appears to be less complex, it is found that the observed compact
H$_2$CO emission can be equally well explained originating from a disk
as from the inner hot core. However, the need for a drastic jump in
abundance depends critically on the properties of the disk. An upper
limit on the para-H$_2$CO abundance in the disk of 
$\sim$4$\times10^{-9}$ is derived adopting a temperature of 100~K, a mass of
0.016~M$_{\odot}$ and a size of 70~AU and assuming that all the flux
on long baselines arises is due to the disk. For \object{IRAS
16293--2422} an upper limit of the abundance in the MM1 disk of 
$\sim$1$\times10^{-9}$ is obtained using a mass of 0.09~M$_{\odot}$ and a
size of 250~AU.   A slightly lower upper limit is obtained for MM2
using a disk mass of 0.12~M$_{\odot}$.  Moreover, the difference
between the morphology of the C$^{18}$O and H$_2$CO emission seems to
indicate that at least the disk around MM1 contains little H$_2$CO,
even though CO may be nearly undepleted. Note that geometrical
effects, for example from the disk potentially shielding parts of the
envelope, can cause differences in the appearance between MM1 and MM2.

H$_2$CO has been detected in protoplanetary disks of T Tauri stars
\citep{Dutrey97, Aikawa03}, where confusion due to an envelope or
outflow is negligible. \citet{Thi04} find an integrated H$_2$CO
$3_{03}\rightarrow2_{02}$ line intensity of 0.14~K\,km\,s$^{-1}$
(about 1.3~Jy\,km\,s$^{-1}$) for the T Tauri (class II) star
\object{LkCa 15} using the IRAM 30~m telescope. The beam-averaged
($10\farcs8$) H$_2$CO abundance is about $4\times10^{-11}$. The OVRO
data for \object{IRAS 16293--2422} presented here give a factor of
about 50 stronger emission. For \object{L1448--C} the emission is
about a factor of 10 stronger after correcting for the distance.
Thus, if the compact H$_2$CO emission were coming from disks, they
would have to be `hotter' or have higher abundances than in the class
II phase.  Accretion shocks in the early stages could be responsible
for such increased disk temperatures. High-angular resolution ($<1
''$) data on the velocity structure of the H$_2$CO lines are needed to
distinguish the disk emission from that of the inner envelope.

\begin{table*}
\caption[]{Predicted H$_2$CO line intensities for \object{L1448--C}}
  \label{predictions}
\begin{tabular}{lcccccccccccccccccc}
\hline\hline
\noalign{\smallskip}
\multicolumn{1}{c}{Transition} &
\multicolumn{1}{c}{Frequency} &
\multicolumn{1}{c}{$E_{\mathrm{u}}$$^a$}  &&
\multicolumn{3}{c}{Envelope$^b$} & &
\multicolumn{3}{c}{Envelope w. jump$^c$} &&
\multicolumn{3}{c}{Envelope w. drop$^d$} &  &
\multicolumn{3}{c}{Envelope+disk$^e$} \\
  &
  \multicolumn{1}{c}{[GHz]} &
  \multicolumn{1}{c}{[K]} & &
  \multicolumn{3}{c}{[K\,km\,s$^{-1}$]} &  &
  \multicolumn{3}{c}{[K\,km\,s$^{-1}$]} &  &
   \multicolumn{3}{c}{[K\,km\,s$^{-1}$]} &  &
  \multicolumn{3}{c}{[K\,km\,s$^{-1}$]} \\
  & & && $0\farcs3$ & $3\arcsec$ & $15\arcsec$ && $0\farcs3$ &  
$3\arcsec$ & $15\arcsec$
  && $0\farcs3$ & $3\arcsec$ & $15\arcsec$
  && $0\farcs3$ & $3\arcsec$ & $15\arcsec$\\

\hline
\noalign{\smallskip}
$3_{03}\rightarrow2_{02}$            & 218.22 & 21   && 20 & 10  & 2.7   && 96 & 11.5 &2.7 && 34 & 6.6 & 2.2 &&137 &12&2.7 \\
$3_{22}\rightarrow2_{21}$            & 218.48 & 68   && 7.2 & 1.8 &0.23  && 109 &3.1 &0.27 && 37 &2.8 &0.22&&160 &3.8&0.30 \\
$5_{05}\rightarrow4_{04}$            & 362.74 & 52   &&  31 & 7.3&0.64   && 135 &8.7 &0.69 &&99 &9.9 &0.66 && 180 &9.4 &0.72 \\
$5_{24}\rightarrow4_{23}$            & 363.95 & 100 && 16 & 1.9&0.12    && 134 &3.5 &0.18 &&76 &4.5 &0.21 && 183 &4.3 &0.21 \\
$5_{42/41}\rightarrow4_{41/40}$ & 364.10 & 241 && 2.4 &0.08&0.004 &&  164 &2.1 &0.08 && 20&0.62 &0.02 && 268 &3.7 &0.15\\
\hline
\noalign {\smallskip}
\end{tabular}

\noindent
$^a$ Energy of the upper energy level involved in the transition.\\
\noindent
$^b$ Constant para-H$_2$CO abundance of $6\times10^{-10}$ throughout the  
envelope.\\
\noindent
$^c$ A jump in abundance of a factor 250, from $6\times10^{-10}$,
when $T>100$~K .\\
\noindent
$^d$ H$_2$CO is depleted by about an order of magnitude over roughly the same region as  
CO ($T_{\mathrm{ev}}=50$~K; $X_{\mathrm{0}}=5\times10^{-9}$; 
$X_{\mathrm{D}}=4.0\times10^{-10}$)\\
\noindent
$^e$ H$_2$CO abundance in the disk is 25 times higher than that in the   
envelope of $6\times10^{-10}$.\\
\end{table*}

\subsection{Predictions for future generation telescopes}
\label{predictions_sec}
In Table~\ref{predictions}, the predicted H$_2$CO line intensities,
integrated over the full extent of the line, are presented for
\object{L1448--C}  using: 1) the envelope model with a constant
abundance of $6\times10^{-10}$, 2) introducing a jump when $T>100$~K
($X_{\mathrm{J}}=1.5\times10^{-7}$;
$X_{\mathrm{D}}=6\times10^{-10}$), 3) a `drop' abundance profile
where H$_2$CO is depleted over the same region as CO
($T_{\mathrm{ev}}=50$~K; $X_{\mathrm{0}}=5\times10^{-9}$;
$X_{\mathrm{D}}=4\times10^{-10}$); and 4) the envelope + disk model
($X_{\mathrm{D}}=6\times10^{-10}$ and disk parameters from
\S\ref{l1448_h2co_sec}).  Beamsizes of $0\farcs3$ and $3\arcsec$ were
assumed to characterize the typical spatial resolutions of current and future
interferometers.  The corresponding intensities picked up by a
single-dish beam of $15\arcsec$ are shown for comparison. Using
single-dish data alone, it will be difficult to discriminate between
these competing scenarios unless the highest frequency lines are
obtained.
 Current interferometers such as OVRO working in the 1~mm window can
however constrain some characteristics of the abundance variations
in the envelopes, such
as the presence of a drop abundance profile.
This seems to be the case for the two
sources studied here, \object{IRAS 16293--2422} and
\object{L1448--C}.

Finally, it is clear that observations at 0.$''$3 will have the
potential to further discriminate a hot core scenario from that of a
warm disk.  Indeed, the much improved resolution and sensitivity of
next generation interferometers such as CARMA, (e)-SMA, (upgraded)
IRAM and ALMA will greatly aid in distinguishing between the competing
scenarios discussed above. They will also provide additional
constraints on the morphology and velocity structure of the gas on
larger and smaller scales.  In addition to searching for outflow
motions, it is of considerable interest to determine if the hot core
gas shows any evidence of infall motions toward the cental
source(s). If chemically processed material such as H$_2$CO and other
organics is in a state of infall toward the central protostars it will
likely be incorporated in the growing protoplanetary disk and become
part of the material from which planetary bodies are formed. With the
present data, it is not possible too uniquely separate infall motions
from those of rotation and/or outflow.

\section{Conclusions}
\label{conclusions}
A detailed analysis of the small scale structure of the two low-mass
protostars \object{IRAS 16293--2422} and \object{L1448--C} has been
carried out.  Interferometric continuum observations indicate that the
inner part of the circumbinary envelope around \object{IRAS
16293--2422} is relatively void of material on scales smaller than the
binary separation ($\sim$5$\arcsec$).  This implies that the clearing
occurs at an early stage of binary evolution and that IRAS 16293--2422
may well develop into a GG Tau-like object in the future. The bulk of
the observed emission for both sources is well described using model
envelope parameters constrained from single-dish observations,
together with unresolved point source emission, presumably due to
circumstellar disk(s).

Simultaneous H$_2$CO line observations indicate the presence of hot
and dense gas close to the peak positions of the continuum emission.
For both \object{IRAS 16293--2422} and \object{L1448--C}, the observed
emission cannot be reproduced with a constant abundance throughout the
envelope.   The H$_2$CO $3_{22}-2_{21}/3_{03}-2_{02}$ ratio on
short baselines ($2-10\arcsec$) is best fit for both sources by an H$_2$CO
`drop' abundance profile in which H$_2$CO, like CO, is depleted (by
more than an order of magnitude) in the cold dense region of the
envelope where $T\lesssim 50$~K, but is relatively undepleted in the
outermost region where $n_{\mathrm{H_2}}<1\times 10^{5}$~cm$^{-3}$.
In the inner region for $T>50$~K, the abundance jumps back to a high
value comparable to that found in the outermost undepleted part.
Additional H$_2$CO abundance jumps ---either in the innermost `hot
core' region or in the compact circumstellar disk--- cannot be firmly
established from the current data set.

Based on the morphology and line widths, little of the observed
emission toward \object{IRAS 16293--2422} is thought to be directly
associated with the known outflow(s).  Instead, the emission seems to
be tracing gas in a rotating disk perpendicular to the large scale
outflow.  For \object{L1448--C}, the morphology of the H$_2$CO line
emission is consistent with the outflow, but the line widths are
significantly smaller and the emission is extended over a large area.
Although the above envelope model with a `drop' abundance profile can
fit the observations, other scenarios cannot be ruled out. These
include the possibility that the outflow (weakly) interacts with the
envelope producing regions of enhanced density and temperature in
which some H$_2$CO is liberated and entrained, and a scenario in which
part of the envelope gas is heated by UV- or X-ray photons escaping
through the outflow cavities.

It is clear that high sensitivity and high spatial resolution
observations are critically needed for a better understanding of the
chemical and dynamical processes operating during star formation. At
the same time, this study also demonstrates that detailed radiative
transfer tools are essential for a correct analysis of such data.  In
the future, the modeling codes need to be extended to multiple dimensions
($>$1D) in order to fully tackle the complex geometry hinted at in
current data sets. Predictions for future arrays are provided that
illustrate their potential to discriminate between competing scenarios
for the origin of the H$_2$CO abundance enhancements, and presumably
that of other complex organics, in low-mass protostars.

\begin{acknowledgements}
C. Ceccarelli and S. Maret are thanked for fruitful discussions on the
interpretation of abundance jumps in low-mass protostars. This
research was supported by the Netherlands Organization for Scientific
Research (NWO) grant 614.041.004, the Netherlands Research School for
Astronomy (NOVA) and a NWO Spinoza grant. FLS further acknowledges
financial support from The Swedish Research Council, and GAB from the
NASA Exobiology program. EvD also thanks the Moore's Scholars program
for an extended visit at the California Institute of Technology.
This paper made use of data obtained at the Owens Valley
Radio Observatory Millimeter Array and the James Clerk Maxwell
Telescope. The authors are grateful to the staff at these facilities
for making the visits scientifically successful as well as enjoyable.
\end{acknowledgements}

\bibliographystyle{aa}

\end{document}